\documentstyle[aps,preprint,tighten,epsfig,amssymb]{revtex}
\begin{document}

\title{The Canonical Transformation and Duality in the 1+1 dimensional 
$\phi^4$ and $\phi^6$ theory}

\author{
Chueng-Ryong Ji$^{a}$
\footnote{e-mail address: ji@ncsu.edu},
Joon-Il Kim$^{b}$
\footnote{e-mail address: jikim@phya.snu.ac.kr},
Dong-Pil Min$^{b}$
\footnote{e-mail address: dpmin@phya.snu.ac.kr},
Andrey V. Vinnikov$^{a,b,c}$
\footnote{e-mail address: vinnikov@phya.snu.ac.kr}}

\address{
\bigskip
$^a$ Department of Physics, North Carolina State University, Raleigh, NC
     27695-8202, USA\\
$^b$ School of Physics and Center for Theoretical Physics,
     Seoul National University, Seoul 151-742, Korea\\
$^c$ Bogoliubov Laboratory of Theoretical Physics, JINR, 141980, 
     Dubna, Russia}

\date{\today}

\maketitle

\begin{abstract}
We investigate the self-organizing nature of relativistic quantum field 
theory in terms of canonical transformation and duality presenting simple
but explicit examples of $(\phi^4)_{1+1}$ and $(\phi^6)_{1+1}$ theories.
Our purpose is fulfilled by applying the oscillator representation (OR) method
which allows us to convert the original strong interaction theory
into a weekly interacting quasiparticle theory that is equivalent
to the original theory.
We discuss advantages of the OR method and compare the results with what 
was already obtained by the method
of Gaussian effective potential (GEP) and the Hartree approximation (HA). 
While we confirm that the GEP results are identical to the Hartree results
for the ground state energy, we found that the OR method 
gives the quasiparticle mass different from the GEP and HA
results. In our examples, the self-organizing nature is revealed by 
the vacuum energy density that gets lowered when the quasiparticles are 
formed.
In the $(\phi^6)_{1+1}$ theory, we found two 
physically meaningful duality-related quasiparticle solutions which have 
different symmetry properties under the transition of quasiparticle 
field $\Phi \rightarrow -\Phi$. However, these two quasiparticle 
solutions yield the identical effective 
potential in the strong coupling limit of the 
original theory.
\end{abstract}



\section{INTRODUCTION}
With the current advances of Relativistic Heavy Ion Collision(RHIC) 
physics, there is a growing interest in discussing
the self-organizing nature of relativistic quantum field theory(RQFT).
The phase transition and the spontaneous symmetry breaking 
anticipated to be observed in the RHIC facilities are the paramount 
examples of the physical phenomena due to the self-organization of 
quantum fields. Since these novel phenomena cannot be easily predicted in 
the ordinary perturbation series, they form highly nontrivial examples
that ought to be analyzed by various nonperturbative methods available
for the RQFTs. 

Although the problem of quantization of interacting fields
is not yet completely solved, it appears that the duality existing in the
RQFTs is the key to handle the strong interaction theories.
According to the duality, the original particle theory of strong couplings 
may be equivalent to the quasi-particle theory of weak couplings. As the 
interactions get stronger, the associate quantum fluctuations get larger 
and consequently the formation of a nontrivial vacuum 
often accompanied by the condensation of fields 
may be energetically more favorable than the adherence to an original trivial 
vacuum without any condensation. The particles moving in the ambient 
nontrivial vacuum condensates may be described by the quasi-particles 
which carry different masses
compare to their original masses defined in the trivial vacuum.
Then, the couplings among the quasi-particles may become weak as the 
couplings among the original particles grow strong. 
When such duality works self-consistently within the RQFT, one may
be able to solve the corresponding strong interaction problem utilizing
the weak coupling developed for the quasi-particles.
For the phenomena involving only weak interactions, the perturbation 
theory may provide systematic predictions and even some physical insight
how the interaction can affect the properties of the system.
Thus, the duality allows us to convert the original nonperturbative 
problem into the quasiparticle perturbative problem.
However, the key issue here is to check if the RQFTs indeed have the
self-organizing characteristics that lead to the duality within 
themselves. To investigate such nontrivial characteristics, one cannot
just rely on the perturbation series from the beginning but need to 
develop a nonperurbative method which may be useful to analyze the 
self-organizing nature of the RQFTs. 

In this work, we discuss such a nonperturbative method that appears to be 
much simpler than other rather well-known methods such as the Hartree 
Approximation (HA) and the Gaussian Effective Potential (GEP) method. 
Following the nomenclature in the literature we call this method as the 
Oscillation Representation (OR) method, although in our view all of these 
nonperturbative methods (OR,GEP,HA) stem from the same principle of
quantum effective action approach\cite{CJT}. 
The OR method was explicitly formulated by Efimov\cite{efimov} and 
was, in fact, used earlier by Chang \cite{chang2} and
Magruder \cite{magruder}, though in some indirect way.
The method is at length described in a monograph \cite{efbook}.
The basic idea of the OR method is to redefine the mass of interacting field
relative to the free one and simultaneously introduce a shift of the field
quantization point leading to a nonzero value of its vacuum condensate. 
This effect is realized in the nature represented by a spontaneous symmetry
breaking mechanism and thus one can expect that it may be a general
feature of quantum field systems with interactions.
However, the OR method has not been utilized as extensively as the methods 
of GEP\cite{ft7,gep,gep4,gep6,mp} and HA\cite{Chang,hartree}. Even in the 
1+1 dimensional scalar field theory such as 
the one we present in this work, not much of the OR results distinguished
from the GEP or HA results have been known or discussed to the best of
our knowledge. 
Although the OR results for the $(\phi^4)_{1+1}$ theory were presented
in Ref.\cite{efbook}, the distinction between the OR and GEP (or HA) 
results for this simple model was not pointed out at length.
Thus, for the present work, we analyze the details of the distinction
in this most simple case first and extend the application to the 
$(\phi^6)_{1+1}$ theory:
\begin{eqnarray}
\label{lagrangian}
{\cal L} = \frac{1}{2} \partial_\mu \phi \partial^\mu \phi
-\frac{1}{2} m^2 \phi^2 - \frac{g}{4} \phi^4 - h \phi^6, 
\end{eqnarray}
where the results from the variational method of the GEP have
also been studied \cite{gep6}.
The GEP is defined as 
\begin{equation}
\label{gep}
{\bar V}_G(\phi_0) = min_\Omega <\phi_0,\Omega|{\cal H}|\Omega,\phi_0>,
\end{equation}
where ${\cal H}$ is the Hamiltonian density corresponding to 
Eq.(\ref{lagrangian}). Here,
$\Omega$ is the mass parameter and $|\Omega,\phi_0>$ is a normalized 
Gaussian wave functional centered on $\phi = \phi_0$, {\it i.e.} 
\begin{equation}
<\phi_0,\Omega|\Omega,\phi_0> = 1, 
<\phi_0,\Omega|\phi|\Omega,\phi_0> = \phi_0.
\end{equation}
As we will discuss the details in the next two sections (Sections II and 
III), the GEP method can in principle give the two OR equations 
that we solve simultaneously in this work. One of them is known
as the ${\bar \Omega}$ equation in the GEP,{\it i.e.}
$\partial <\phi_0,\Omega|{\cal H}|\Omega,\phi_0>/\partial \Omega = 0$
at $\Omega = {\bar \Omega}$, and the other OR equation is equivalent
to $d{\bar V}_G(\phi_0)/{d \phi_0} = 0$. However, in the GEP method,
these two equations are not solved simultaneously in practice.
It turns out that the quasiparticle mass defined by $d^2 {\bar 
V}_G(\phi_0)/{d \phi_0^2}$ in the GEP is not same with ${\bar \Omega}$
which is the simultaneous solution of the quasiparticle mass from the two 
OR equations.
In this paper, we therefore pay more
attention to the distinct aspects of the OR results compared to the GEP 
and 
HA results and discuss the property of the OR results which was not 
mentioned in the more great detail before.

Formally, the OR is settled by requiring that the Hamiltonian operator
should be written in terms of creation and annihilation operators
of an oscillator basis with an appropriate frequency and in the correct
form defined by\\
(1) all field operators in the total Hamiltonian $H = H_0 +H_I$ are
written in the normal ordered product,\\
(2) the Hamiltonian $H_0$ is quadratic over the field operators,\\
(3) the interaction Hamiltonian $H_I$ contains field operators in powers
more than two.\\
Using this method with the cannonical transformation of quantum fields, 
one may allow to build a standard perturbation theory 
for the quasiparticles in the region of the coupling constants where
the original theory is nonperturbative. Thus, we utilize the OR method 
to transform the original theory which is intrinsically nonperturbative
to the equivalent theory of the quasiparticle
representation that can be solved rather easily by the perturbation 
method. 

As we will discuss, the exact OR results include several  
vacuum solutions yielding different mass values of the quantum 
field for fixed values of coupling constants.
Among those multiple solutions, however, it is certainly not difficult 
to select the non-trivial solutions that are physically meaningful 
because of the duality consideration. In this work, we will thus focus 
only those duality-related solutions which are not hindered from the 
physical interpretation.
Nevertheless, we note that the OR method generates also the solutions that
cannot be interpreted as straightforward as the ones we focus in this 
paper. In particular, we find a class of solution that doesn't exhibit 
the anticipated duality property and another class of 
solution that we call spurious in the sense that it yields
the pure imaginary value for the vacuum condensation.
We do not yet know all the physical meaning of these classes of solutions,
but the generation of multiple solutions are expected because the OR 
equations are the nonlinear algebraic equations relating the quasiparticle
mass and the field condensation with the parameter of coupling constant  
as we will present in details for the rest of the paper.

In Section II, we present more details of the OR method using the 
$\phi^4$ theory which can be easily deduced from the Lagrangian 
density of $\phi^6$ theory by taking $h=0$. We compare the OR results 
with the results obtained by the other available methods such as the GEP 
and the HA as well as the available lattice result.
In Section III, we apply the OR method to 
the $(\phi^6)_{1+1}$ theory and discuss the effects from the presence of 
$\phi^6$ interaction to the duality-related vacuum solutions.
We analyze the quasiparticle mass and the energy density in
the entire domain of coupling parameter space and 
present the classical potential for the nontrivial solutions.
We discuss two physically meaningful duality-related quasiparticle 
solutions with different symmetry properties and show that they however
generate the identical effective potential in the strong coupling
limit of the original theory. 
Summary and conclusions follow in Section IV. In the Appendix A, 
we briefly summarize some details of the HA in the
$(\phi^4)_{1+1}$ theory and discuss an equivalence between the two 
nontrivial vacuum solutions in this approximation. In the Appendix B,
we present a proof of unitary inequivalence between the bifurcated 
nontrivial solutions from the OR method in the infinite volume limit.
In the Appendix C, we summarize the results of the spurious solutions
with some discussion.
 
\section{THE OSCILLATOR REPRESENTATION (OR) METHOD}
The usual recipe to produce a quantum system comes from the correspondence
principle. Namely, a quantum system can be obtained from
its classical analog by assuming that the field is not a $c$-number
but an operator satisfying relevant commutation relations.
The oscillator representation method is a generalization
of this scheme. To start with, let us consider the $\phi^4$ theory
as a simple example.
 
The Hamiltonian density of the classical field is given by
\begin{equation}
{\cal H} = \frac{1}{2}\pi^2+\frac{1}{2}(\nabla \phi)^2+\frac{1}{2}
m^2\phi^2+\frac{g}{4}\phi^4.
\end{equation}
To plug the quantum physics in, we postulate that the field $\phi$ and 
the corresponding conjugate momentum $\pi$ are operators which satisfy 
canonical equal time commutation relations
\begin{equation}
[\phi(x),\pi(y)]_{x_0=y_0}= i\delta({\bf x}-{\bf y}).
\label{commut}
\end{equation}

The usual approach to quantize the field, {\it i.e.} to find the operators 
$\phi$ and $\pi$ which satisfy the condition (\ref{commut}) is to start from 
the free field, where $g$=0. Then the operators are
\begin{eqnarray}
\phi_m(x) = \int \frac{d {\bf k}}{\sqrt{2\pi}} \frac
{1}{\sqrt{2\omega_m({\bf k})}}
(a_m({\bf k})e^{ik\cdot x} + a_m^+({\bf k})e^{-ik\cdot x}), \nonumber \\
\pi_m(x) = \frac{1}{i}\int \frac{d {\bf k}}{\sqrt{2\pi}}
{\sqrt{\frac{\omega_m({\bf k})}{2}}}(a_m({\bf k})e^{ik\cdot x} - a_m^+({\bf k})
e^{-ik\cdot x}),
\label{fieldop}
\end{eqnarray}
where $k=(\omega_m,{\bf k})$, $\omega_m=\sqrt{{\bf k}^2+m^2}$ and the
operators of creation and annihilation of the particle of mass $m$ and
momentum ${\bf k}$ satisfy the commutation relations
\begin{eqnarray}
[ a_m({\bf k}),a_m^+({\bf k'})] = \delta({\bf k} -{\bf k'}), \nonumber \\
{[}a_m({\bf k}),a_m({\bf k'}) ] = [ a_m^+({\bf k}),a_m^+({\bf k'})] =0.
\label{operat}
\end{eqnarray}
With this definition the state $a_m^+({\bf k})|0\rangle$ is an eigenstate
of both
free Hamiltonian and momentum operators with corresponding eigenvalues
$\omega_m$ and ${\bf k}$.

In the case of interacting fields the states $a_m^+({\bf k})|0\rangle$ are
not
already eigenstates of the full Hamiltonian. Nevertheless,
in the case of small interaction strength, the free representation
of field operators can be a good starting point to develop a perturbation 
theory. The Hamiltonian density operator is then written as
\begin{equation}
{\cal H}\rightarrow {\cal H}_m = \frac{1}{2}\pi^2_m+
\frac{1}{2}(\nabla \phi_m)^2+\frac{1}{2}m^2\phi^2_m+\frac{g}{4}\phi_m^4.
\label{quantham}
\end{equation}

Now, we use the canonical transformation of the field operators
to build the quasiparticle representation of the same Hamiltonian.
We consider the essential point of the reparameterizations,
{\it i.e.} the change of mass, which can affect the choice of 
representation for the free field quantization.
For the free field operators, the change of mass $m\rightarrow M$ of 
particles is described by the Bogoliubov-Valatin transformation:
\begin{eqnarray}
a_M({\bf k})=\frac{1}{2}\left( \sqrt{\frac{\omega_m({\bf k})}
{\omega_M({\bf k})}} + \sqrt{\frac{\omega_M({\bf k})}
{\omega_m({\bf k})}} \right) a_m({\bf k})-
\frac{1}{2}\left( \sqrt{\frac{\omega_m({\bf k})}
{\omega_M({\bf k})}} - \sqrt{\frac{\omega_M({\bf k})}
{\omega_m({\bf k})}} \right) a^+_m({\bf -k}), \nonumber \\
a^+_M({\bf k})=\frac{1}{2}\left( \sqrt{\frac{\omega_m({\bf k})}
{\omega_M({\bf k})}} + \sqrt{\frac{\omega_M({\bf k})}
{\omega_m({\bf k})}} \right) a^+_m({\bf k})-
\frac{1}{2}\left( \sqrt{\frac{\omega_m({\bf k})}
{\omega_M({\bf k})}} - \sqrt{\frac{\omega_M({\bf k})}
{\omega_m({\bf k})}} \right) a_m({\bf -k}).
\end{eqnarray}
Here, the free field operators with mass $M$ are expressed in terms of 
the field operators with mass $m$.

A consequence of the ambiguity of the mass definition due to interactions
comes from the ambiguity in the choice of the initial representation of the 
interacting field. This was emphasized in Coleman's paper \cite{coleman}.
He also proposed a useful technique on how to redefine a normal 
ordered product of any number of field operators with respect to a new 
value of mass in the case of (1+1) scalar field theory. The redefinition 
after the mass change is given by the formulas \cite{chang2,coleman}:
\begin{equation}
N_m\left(\frac{1}{2}\pi^2_m+\frac{1}{2}(\nabla\phi_m)^2\right)=
N_M\left(\frac{1}{2}\pi^2_M+\frac{1}{2}(\nabla\phi_M)^2\right)+
\frac{1}{8\pi}(M^2-m^2),
\label{normfree}
\end{equation}
\begin{equation}
N_m(e^{\imath \beta\phi_m})=\Bigl ( \frac{M^2}{m^2} {\Bigr )}^{\beta^2/8\pi}
N_M(e^{\imath \beta\phi_M}),
\label{normexp}
\end{equation}
where $\beta$ is some arbitrary $c$-number, $N_m$($N_M$) stands for 
the normal ordering of an operator with respect to
the initial mass $m$ (the new mass $M$), $\phi_m$
is the free quantum field defined by Eq.(\ref{fieldop}) and $\phi_M$ is
the quantum field of independent quasi-particles for which 
the representation
given by Eq.(\ref{fieldop}) is valid after changing $m\rightarrow M$.
The correspondence between the canonical transformation and the 
renormalization group equation is discussed in Ref.\cite{efbook}.
The expression (\ref{normexp}) can be rewritten as
\begin{equation}
N_m(\phi^n_m)=n!\sum\limits^{[\frac{n}{2}]}_{j=0} \Bigl ( \frac{1}{8\pi}
\ln(\frac{M^2}{m^2}) \Bigr ) ^j\frac{(-1)^j}{j!(n-2j)!}N_M(\phi^{n-2j}_M),
\label{normord}
\end{equation}
where $\bigl [ \frac{n}{2} \bigr ] $ is the integer part of $\frac{n}{2}$.

The most simple illustration of nontrivial solutions in this case is the 1+1
dimensional $\phi^4$ theory. Rewriting the Hamiltonian (\ref{quantham})
in terms of the quasi-particles using Coleman's formula of normal ordering
rearrangement, one gets
\begin{eqnarray}
N_m({\cal H}_m) &=& N_M({\cal H}_M) \nonumber \\
&+&\frac{1}{2}(m^2-M^2) N_M(\phi^2_M) 
+\frac{1}{8\pi}(M^2-m^2) -\frac{m^2 t}{8\pi}
-\frac{3gt}{8\pi} N_M(\phi^2_M)+\frac{3gt^2}{64\pi^2},
\label{phim}
\end{eqnarray}
where $t=\ln\bigl ( \frac{M^2}{m^2} \bigr ) $.

For a consistent
description of the dynamics in terms of the quasi-particles one
should provide the right form of the Hamiltonian, {\it i.e.} the free part 
should correspond to that of the field with mass $M$ and the 
interacting part should contain
only terms with powers of $\phi_M$ larger than 2 \cite{efbook}. As it can
be seen, the expression (\ref{phim}) itself can not provide this requirement
for the field with mass other than $m$. Revising just the mass does not
lead to a nontrivial solution and thus an additional transformation 
is obviously required. The latter can be done by shifting the quantization 
point of the field, which makes the quanta
of the field to be produced around some vacuum expectation value 
$\langle \phi \rangle \ne 0$.

Thus, let us now change the field variable from $\phi_M(x)$ to $\Phi_M(x)$
for the inclusion of the field condensation denoted by $b (=\langle 
\phi_M \rangle)$:{\it i.e.}, 
\begin{equation}
\phi_M(x) = \Phi_M(x) + b.
\end{equation}
Then the Hamiltonian density is written as \cite{efbook}:
\begin{equation}
N_M({\cal H})=N_M({\cal H}_M^{right})+{\cal H}_1+\varepsilon_M,
\end{equation}
where ${\cal H}_M^{right}$ is the right form of Hamiltonian density for
the field with mass $M$,
\begin{equation}
{\cal H}_M^{right}=\frac{1}{2}\pi^2_M + \frac{1}{2}(\nabla\Phi_M)^2 +
\frac{M^2}{2}\Phi^2_M + \frac{g}{4}\Phi^4_M  +
g b\Phi^3_M,
\label{phi4ham}
\end{equation}
$\varepsilon_M$ is the energy density of the vacuum for the quasiparticle 
of mass $M$ up to the zeroth
order of the dimensionless effective coupling $g/M^2$,
\begin{equation}
\varepsilon_M=\frac{m^2b^2}{2}+\frac{gb^4}{4}-\frac{m^2 t}{8\pi}-\frac{3gb^2 t}{8\pi}
+\frac{3gt^2}{64\pi^2}+\frac{M^2-m^2}{8\pi},
\label{energy}
\end{equation}
and
\begin{equation}
{\cal H}_1 = \Bigl \{ \frac{1}{2}(m^2-M^2)+ \frac{3}{2}g b^2 -\frac{3gt}{8\pi}
\Bigr \}  \Phi^2_M + \Bigl \{m^2 b +g b^3 -\frac{3g b t}{4\pi}\Bigr \} \Phi_M
\end{equation}
is a remainder. To have the Hamiltonian in the right form the remainder
must be equal to 0. Therefore, we have two equations of two variables, 
$t$ and $b$, as follows:
\begin{equation}
\frac{1}{2}(1-\frac{M^2}{m^2})+ \frac{3}{2}G b^2 -\frac{3Gt}{8\pi} = 0,
\label{OR1}
\end{equation}
and
\begin{equation}
b \Bigl (1  +G b^2 -\frac{3G t}{4\pi} \Bigr ) = 0,
\label{OR2}
\end{equation}
where $G$ is the dimensionless coupling of the initial theory, {\it i.e.}
$G=g/{m^2}$.

We note here the correspondence to the GEP discussed in Section I.
The energy density $\varepsilon_M$ given by Eq.(\ref{energy}) corresponds
to $<\phi_0,\Omega|{\cal H}|\Omega,\phi_0>$ presented in Eq.(\ref{gep}).
Minimizing $<\phi_0,\Omega|{\cal H}|\Omega,\phi_0>$ (or $\varepsilon_M$)
with respect to $\Omega$ (or $t$), one gets the GEP given by 
Eq.(\ref{gep}) with the so-called ${\bar \Omega}$ equation
\footnote{ See 
Eq.(4.12) of Ref.\cite{gep4} which 
was rederived as Eq.(2.16a) of Ref.\cite{mp} using Bogoliubov 
transformations.}
which is identical to Eq.(\ref{OR1}) that is one of the OR equations. 
The other OR equation, Eq.(\ref{OR2}), can be obtained by making the 
total (not partial) derivative of the GEP ({\it i.e.}, Eq.(4.11) of 
Ref.\cite{gep4} or Eq.(2.16b) of Ref.\cite{mp}) with respect to the 
condensation $\phi_0 (= b)$ and setting it to zero, 
{\it i.e.} $d{\bar V}_G(\phi_0)/{d \phi_0} = 0$. 
One should note, however, that the GEP is a function of 
the global field condensate $\phi_0$ not the local field $\phi(x)$. 
Thus, the GEP is not a part of interaction Hamiltonian describing the 
field dynamics.
In our work, instead of using the GEP,
we will directly use the interaction Hamiltonian and the corresponding
classical potential (see {\it e.g.} Eq.(\ref{Vclass})).

The OR equations (Eqs.(\ref{OR1}) and (\ref{OR2})) have three solutions
and one of them is trivial: $t=0$, $b=0$. The two nontrivial mass 
solutions 
are shown in Fig.1 as solid and dashed lines. 
The dashed line solution on Fig.1 becomes similar to the trivial one when 
the dimensionless coupling becomes large: $M_{G\to\infty}\to m$,
$b_{G\to\infty}\to 0$. However, the solid line solution from Fig.1 is what 
we are interested in because it fulfills the duality. For this case,
$M_{G\to\infty}^2\to {\frac {3}{2\pi}}m^2 G\ln G$,
$b_{G\to\infty}\to \sqrt{\frac{3}{4\pi}\ln G}$.
As one can see in this solution, when the dimensionless coupling $G$ of 
the initial theory becomes large, the dimensionless couplings of the 
quasiparticle theory, $\chi^{(3)}=gb/M^2$ and $\chi^{(4)}=g/M^2$,
get small\cite{efbook}: $\chi^{(3)}_{G\to\infty}\to 
\sqrt{\frac{\pi}{3\ln G}}$,
$\chi^{(4)}_{G\to\infty}\to \frac{2\pi}{3 G\ln G}$. Thus, the
quasiparticle theory becomes perturbative when the initial coupling $G$
gets very strong.
The physical mass in this regime is close to
$M\approx m \sqrt{{\frac{3}{2\pi}}G \ln G}$, and the corrections are in
the order of ${\cal O}\bigl((\chi^{(3)})^2\bigr)$. In this way, we can 
see the duality between the initial nonperturbative theory of unbroken 
$\phi \rightarrow -\phi$ symmetry with a large dimensionless coupling $G$ 
and the quasiparticle theory of broken $\Phi \rightarrow -\Phi$ symmetry
with small couplings $\chi^{(3)}$ and $\chi^{(4)}$.
This duality is exact because the canonical transformation leaves 
the Hamiltonian intact but changes only its 
representation in terms of the quasiparticle field variable $\Phi_M(x)$. 

As we discussed earlier, the multiple solutions are not unexpected because
the OR equations, Eqs.(\ref{OR1}) and (\ref{OR2}), are algebraically
nonlinear. However, it can be shown (see Appendix B) that the two vacua 
associated with different values of $t$ or $b$ are unitarily inequivalent 
in the infinite volume limit. Thus, the two nontrivial solutions are not 
connected to each other. 
The bifurcation of quasi-particle masses appear above the coupling 
$G \approx 9.04$.
However, the correct critical coupling of the phase transition should be
determined from the comparison of the energy density between the trivial
and condensed vacua. In Fig.2, we present the energy density of the 
nontrivial solutions and find that the condensed vacuum energy density
of the duality-related solution gets lower than the trivial vacuum energy 
density only above $G \approx 10.21$. This value $G \approx 10.21$ 
coincides with the critical coupling constant obtained by the GEP method
since our energy density given by Eq.(\ref{energy}) is equal to the GEP
at the minimum. This value is remarkably close to the 
critical coupling constant (10.24(3)) obtained by the lattice 
calculation\cite{lattice}.
Nevertheless, we note that our OR result doesn't take into account the
the higher order corrections in the quasiparticle effective coupling 
$g/M^2$ and obtain the first order phase transition  
instead of the second order phase transition which may be more 
accurate description of the phase transition in $(\phi^4)_{1+1}$ theory
\cite{Chang,simon}. In this work, we thus focus only on the region where 
the original coupling is much larger than the critical coupling so that the 
perturbative calculation in terms of quasiparticle degrees of freedom is 
allowed.      

Although there is a coincidence of the critical coupling,
it is not clear if the values of quasi-particle masses defined in the GEP, 
HA and OR methods also coincide. In the GEP approach, it seems only
natural to define the quasiparticle mass as the second derivative 
of the GEP\footnote{ See for example Eq.(3.2) of Ref.\cite{gep4}.}
, {\it i.e.} $d^2 {\bar V}_G(\phi_0)/d \phi_0^2$.
As discussed earlier, in the GEP approach a variational method is taken 
to minimize the GEP satisfying just Eq.(\ref{OR1}). 
Confirming the well-known equivalence between the GEP method and 
the HA for the ground-state energy \cite{CJT,ft7},
the GEP result using this definition of 
quasiparticle mass, {\it i.e.} the second derivative of the GEP, indeed 
agrees precisely with the Hartree results (squares and circles in 
Fig.\ref{geff})\footnote{In Chang's paper \cite{Chang} he found two 
different types of nontrivial solutions using Hartree method ($4\pi c^2 
>3$ and $4\pi c^2 <3$ in his notations). Nevertheless, the two type of 
solutions lead in fact to
the same physical results as shown in Fig.\ref{geff}. 
For more details, see the Appendix A.}.
However, one should note that the second 
derivative of the GEP at the minimum $\phi_0 = b$ does not yield the mass 
$M$ of the quasi-particle precisely but somewhat modified value 
$M\sqrt{\frac{4\pi M^2 - 6 g}{4\pi M^2 + 3 g}}$; {\it i.e.}
with the notations used in Eq.(4.11) of Ref.\cite{gep4}, the second 
derivative of the GEP, ${\bar V}_G$, is given by
\begin{equation}
\label{gepdd}
\frac{d^2{\bar V}_G}{d\phi_0^2}\bigg|_{\phi_0 = \pm 
\sqrt{\frac{x}{8 {\hat \lambda_B}}}} = x \Bigl ( \frac{\pi x - 6 {\hat 
\lambda_B}}{\pi x + 3 {\hat \lambda_B}} \Bigr ) \neq x, 
\end{equation}
where $x = \frac{M^2}{m^2}$, ${\hat \lambda_B} = \frac{g}{4 m^2}$ and 
$\phi_0 = b$ in our notations. Removing $t (= \ln x)$ in Eqs.(\ref{OR1}) 
and (\ref{OR2}), one can immediately get $b = \pm \frac{M}{\sqrt{2 g}}$ 
identical to $\phi_0 = \pm \sqrt{\frac{x}{8 {\hat \lambda_B}}}$.
This is the reason why the GEP result given by Eq.(\ref{gepdd}) 
shown by dotted line in Fig.\ref{geff} 
doesn't  coincide either of the bifurcated OR solutions (solid and dashed 
lines). Of course, if both the ${\bar \Omega}$ equation and the GEP 
minimum condition $d {\bar V}_G(\phi_0)/d \phi_0 = 0$ are solved 
simultaneously
to determine ${\bar \Omega}$, then it is identical to the OR method and
the value of ${\bar \Omega}$ coincides with the quasiparticle mass $M$
defined in the OR method. However, we note once again that the GEP and HA 
methods are quite different from the OR procedure because the GEP and HA 
methods rely on the variational procedure while the OR method is 
explicitly solving both Eqs.(\ref{OR1}) and (\ref{OR2}) simultaneously. 

As we have already discussed, the solid line in Fig.\ref{geff} 
exhibits a duality between the theory with mass $m$ based on the trivial 
vacuum $b=0$ and the theory with mass $M$ based on the nontrivial vacuum 
$b \neq 0$. Namely, a strong coupling theory with mass $m$ is identical 
to a weak coupling theory with mass $M$. A similar observation of duality 
can be found in the solutions obtained by GEP and HA (see also Appendix 
A), although the quasiparticle masses defined in the GEP and HA approaches
do not coincide with the quasiparticle mass defined in the OR method.
We also note that the GEP result (dotted line) approaches to
the OR result given by the solid line in the limit $G \to \infty$
since the terms of $x$ overwhelm the terms of $\hat \lambda_B$ in 
Eq.(\ref{gepdd}) for that limit. 

In Ref. \cite{mp}, it was argued that the method of
Bogoliubov-Valatin transformations gives the same results as the GEP.
In fact all of the nonperturbative methods that we discuss in this 
work,{\it i.e.} OR, GEP and HA, can be described formally in terms of 
Bogoliubov-Valatin transformations because all of them use only the 
change of particle mass and the field shift to get the new representation 
of the field variables.
However, these transformations contain parameters (mass ratio
and the value of the shift) which can not be determined from
the transformations themselves. To fix the parameters, some
additional requirement is necessary. 
The authors of Ref.\cite{mp} used essentially the same variational 
requirement as the GEP
and therefore it is very natural that they found the same results as the
GEP. On the other hand, in the OR method, we are not using the GEP at all
but relying on the form of the Hamiltonian described in Section I.

Applying the correspondence principle to the Hamiltonian (\ref{phi4ham}),
we now consider the classical potential given by
\begin{equation}
V^{class}(\Phi_M^{class}=\phi_M^{class}-b)=\frac{M^2}{2}(\phi^{class}_M-b)^2 
+\frac{g}{4}(\phi^{class}_M-b)^4  + g b(\phi^{class}_M-b)^3,
\label{Vclass}
\end{equation}
which is written here in terms of the classical field 
corresponding to the initial, unshifted field $\phi_M$.
As it can be easily seen, the potential has extremum at $\phi^{class}_M=b$.
This is provided by the demand not to have terms in Hamiltonian which
are linear in $\Phi_M$.
The double derivative in this point is larger than 0 and equal to
$M^2$,
{\it i.e.} the classical potential has minimum at $\phi^{class}_M=b$ and
the mass corresponding to oscillations around the minimum is precisely $M$.
That is why the OR method provides self-consistent procedure
of quantization of interacting fields and interpretation of the
solutions as particle excitations is valid. (In contrast, 
$\frac{d^2{\bar V}_G}{d \phi_0^2}|_{\phi_0 = b} \neq \frac{M^2}{m^2}$ for 
the GEP as shown in Eq.(\ref{gepdd}).) In addition, one can easily 
prove the symmetry of $V^{class}$ under the 
exchange of $\phi^{class}_M$ and 
$-\phi^{class}_M$,{\it i.e.}$V^{class}(\phi_M^{class}) = 
V^{class}(-\phi_M^{class})$, 
utilizing the OR equations (Eqs.(\ref{OR1}) and (\ref{OR2})). 

In Fig.\ref{g4new}, the effective potential $V^{class}$ given by 
Eq.(\ref{Vclass}) is presented. 
As one can see in Fig.\ref{g4new}, 
the bifurcated solutions for $G>9.04$ indeed generate two 
effective potentials (I,II) at a given coupling $G$, {\it e.g.} $G =$ 10 
or 
12. Both potentials (I and II) in Fig.\ref{g4new} have clear minima and the 
second derivative at each minimum generates the corresponding quasi-particle 
mass $M$ precisely. The growth of bifurcation is also evident from the 
comparison of the potentials I and II for $G =$ 10 and 12. 
We note however that only the potential I provides the physically 
meaningful duality-related solution and the potential II doesn't 
satisfy the duality that we discuss in this work.

An analysis of realistic models using the oscillator representation method
is more complicate (see, for example, discussion for 3+1 dimensional
$\phi^4$ theory in \cite{efbook}) due to substantial sophistication of 
the renormalization procedure. Even the simple models however were not
fully studied in the framework of OR method. 
As presented in this Section, the comparison of the OR predictions with 
the GEP and Hartree results has been done for
1+1 dimensional $\phi^4$ theory \cite{gep,Chang}. In higher
dimensions, the variational methods such as the GEP and HA may have a 
rather fundamental difficulty. 
The reason of this problem originates from the 
error estimates of the approximation
in the GEP and Hartree methods.
For instance, the HA replaces the interaction term $\lambda \phi^n$ for 
$n>2$ by the kinetic term $\Delta m^2 \phi^2+E_0$. The task of getting 
the better approximation is to minimize the 
positively defined measure $||\lambda \phi^n -
\Delta m^2\phi^2-E_0||$. Nevertheless, the vacuum
expectation value of the operator $\lambda \phi^n - \Delta m^2\phi^2-E_0 $
is not positively defined. Thus, it is necessary
to calculate the vacuum expectation value for its square that
is always positive. In that case, even if the interaction term itself
is renormalizable, its square may be not (in particular, for $D>2$).
That is why in higher dimensions the GEP and Hartree methods come across
a renormalization problem in computing their error measures\cite{Munoz}.
More discussion about the difficulty in the GEP method for the higher 
dimension can be found in Ref.\cite{wudka}.
However, this is not the case for the OR method, since it does not use any 
variational procedure.
 
Therefore, among the three (OR,GEP,Hartree), the OR method seems to be 
most attractive. The reasons may be summarized as follows:\\ 
i) The problems of renormalization in higher dimensions are technical, 
but not principal.\\ 
ii) The Hamiltonian is identically rewritten via canonical transformation
and consequently the transformed Hamiltonian keeps the identical 
information as the original Hamiltonian.\\
iii) The second derivative of effective potential gives the 
quasi-particle mass precisely in the OR, while the 
others give somewhat modified values.

\section{Application of the OR method to the $(\phi^6)_{1+1}$ theory}
Using the OR method, we now compute the parameters of different phases 
for the scalar field theory with both $\phi^4$ and $\phi^6$ interactions. 
The starting classical Hamiltonian density is
\begin{equation}
{\cal H}_m=\frac{1}{2}\pi^2_m+\frac{1}{2}(\nabla\phi_m)^2
+\frac{m^2}{2}\phi^2_m
+\frac{g}{4}\phi^4_m + h\phi^6_m.
\label{clasphi6}
\end{equation}
The way to proceed is the same as in the case of $\phi^4$ theory. 
Using the formulas (\ref{normfree}) and (\ref{normexp}) and shifting the 
quantization point, we get:
\begin{equation}
N_m({\cal H}_m)=N_M \bigl ({\cal H}_M^{right} + {\cal H}_1 \bigr ) +
\varepsilon_M,
\label{hamilt6}
\end{equation}
where the ``right'' Hamiltonian density ${\cal H}_M^{right}$ is given by
\begin{eqnarray}
\label{rightH}
{\cal H}_M^{right} &=& \frac{1}{2}\pi^2_M+\frac{1}{2}(\nabla\Phi_M)^2+
\frac{M^2}{2}\Phi_M^2 + \frac{g}{4}\Phi_M^4 +h\Phi_M^6 + gb\Phi_M^3 
\\ \nonumber &+& 6hb\Phi_M^5+15hb^2\Phi_M^4
+ 20hb^3\Phi_M^3-\frac{15ht}{4\pi}\Phi_M^4-\frac{15hbt}{\pi}\Phi_M^3,
\end{eqnarray}
the remainder ${\cal H}_1$ is given by
\begin{eqnarray}
{\cal H}_1 &=& \Bigl ( \frac{1}{2}(m^2-M^2)+\frac{3gb^2}{2}-\frac{3gt}{8\pi}
+15b^4h-\frac{45b^2ht}{2\pi}+\frac{45ht^2}{16\pi^2}\Bigr )\Phi_M^2 
\nonumber \\
&+&\Bigl ( m^2 b+gb^3-\frac{3gbt}{4\pi}+6hb^5-\frac{15hb^3t}{\pi}+
\frac{45bht^2}{8\pi^2} \Bigr )\Phi_M,
\end{eqnarray}
and the energy density $\varepsilon_M$ of the vacuum is given by
\begin{eqnarray}
\varepsilon_M 
&=&\frac{m^2b^2}{2}+\frac{gb^4}{4}-\frac{m^2t}{8\pi}-\frac{3gb^2t}
{8\pi}+\frac{3gt^2}{64\pi^2} \nonumber \\
&+&\frac{1}{8\pi}(M^2-m^2)+hb^6-\frac{15hb^4t}{4\pi}
+\frac{45hb^2t^2}{16\pi^2}-\frac{15ht^3}{64\pi^3}.
\end{eqnarray}
As in the last section, here we use the notation $\Phi_M = \phi_M - b$.

The requirement to have the Hamiltonian in the right form means that
the remainder ${\cal H}_1$ is equal to zero. This gives the following two 
equations for $t$ and $B$ ($B=b^2$):
\begin{equation}
1-e^t+3GB-\frac{3Gt}{4\pi}+30HB^2-\frac{45HBt}{\pi}+\frac{45Ht^2}{8\pi^2}=0,
\label{equ1}
\end{equation}
\begin{equation}
1+GB-\frac{3Gt}{4\pi}+6HB^2-\frac{15HBt}{\pi}+\frac{45Ht^2}{8\pi^2}=0,
\label{equ2}
\end{equation}
where
$G=g/m^2$ and $H=h/m^2$ are the dimensionless couplings.
The conditions (\ref{equ1}) and (\ref{equ2}) again provide the minimum of 
the classical potential at $\phi^{class}_M=b$ and the right value of the
double derivative around it.

Solving Eqs. (\ref{equ1}) and (\ref{equ2}) numerically, we found
five nontrivial mass solutions shown in 
Figs. \ref{phasesnew1}, \ref{phasesnew2} and \ref{phasesnew3}. The 
corresponding vacuum energy densities are presented in Figs. \ref{vac6sym},
\ref{vac6brok} and \ref{vac6spur}.
In Fig.\ref{vac6sym}, two identical plots are shown in two different 
angles of view to make clear presentation of the result.
Among the five, two solutions correspond to the symmetric phase,
{\it i.e.} $b=0$. They are represented by the upper and lower branches
(or sheets) of the manifold shown in Fig.\ref{phasesnew1}. The vacuum 
energy densities for these solutions are shown in Fig.\ref{vac6sym}.
The upper and lower branches in Fig.\ref{phasesnew1} correspond to the
lower and upper branches in Fig.\ref{vac6sym}, respectively.
The upper branch in Fig.\ref{phasesnew1} is the physically meaningful
duality-related solution while the lower branch in Fig.\ref{phasesnew1} 
does not provide the duality that we are interested in this work. The 
duality-related solution
has the lower energy density as shown in Fig.\ref{vac6sym}.
The dimensionless couplings which correspond
to the $\Phi^6$ and $\Phi^4$ interactions of the duality-related
quasiparticle theory are given by
\begin{equation}
\chi_s^{(6)} = \frac{h}{M_s^2},~~~\chi_s^{(4)}=\frac{g}{4M_s^2}-
\frac{15ht}{4\pi M_s^2}.
\label{pertunbr}
\end{equation}
Note here that $\chi_s^{(4)}$ includes both couplings of $g$ and $h$. This is
due to the self-organization of fields via the quantum fluctuations in the
vacuum, {\it i.e.} the original $\phi^6$ interaction induces the effective
$\Phi^4$ interaction with the negative effective coupling in this case.
Their dependencies on $H$ are presented in Fig.\ref{coup1},
where the original $\phi^4$ interaction coupling $g$ is put to zero for
simplicity. It can be seen from Fig.\ref{coup1}, both $\chi_s^{(4)}$
and $\chi_s^{(6)}$ become small as $H$ grows.
Thus, the qusiparticle theory in the domain of large $H$ can be solved by
a standard perturbation method.

Among the rest three solutions, the two solutions corresponding to the
broken-symmetry phase with the nonzero real condensation ($b=\sqrt{B}$) 
are shown as the two branches in Fig.\ref{phasesnew2}. The vacuum energy
densities for these solutions are shown in Fig.\ref{vac6brok}.
Again the upper (lower) branch of Fig.\ref{phasesnew2} corresponds to
the lower (upper) branch of Fig.\ref{vac6brok}.
Similar to the case of 
symmetric phase (Fig.\ref{phasesnew1}), the lower branch in 
Fig.\ref{phasesnew2} does not provide the duality that we are interested
in this work. However, the upper branch in Fig.\ref{phasesnew2} corresponds
to the duality-related solution and gives the lower value of the vacuum
energy density (the lower branch in Fig.\ref{vac6brok}). 
For this broken-symmetry phase, the dimensionless couplings of the 
quasiparticle interactions are given by
\begin{eqnarray}
\chi_{bs}^{(6)}=\frac{h}{M_{bs}^2},~~~
\chi_{bs}^{(5)}=\frac{6hb}{M_{bs}^2},~~~
\chi_{bs}^{(4)}=\frac{g}{4M_{bs}^2}+
\frac{15hb^2}{M_{bs}^2}-\frac{15ht}{4\pi M_{bs}^2},
\nonumber \\
\chi_{bs}^{(3)}=\frac{gb}{M_{bs}^2}+\frac{20hb^3}{M_{bs}^2}-
\frac{15hb t}{\pi M_{bs}^2}.
\label{pertbr}
\end{eqnarray}
The $H$-dependence of these couplings is shown in Fig.\ref{coup2}
(again, we put $g=0$ to simplify the presentation).
At sufficiently large $H$ the couplings get small and the quasiparticle 
theory becomes perturbative. When the dimensionless
couplings of the symmetric solution go to zero, 
$\chi^{(6)}_s,\chi^{(4)}_s \to 0$,
the dimensionless couplings of the broken-symmetry theory also go to zero
as shown in Fig.\ref{couplings}.

The last solution shown in Fig.\ref{phasesnew3} has a pure imaginary value
for the condensate $b$ ({\it i.e.} negative $B$). In Fig.\ref{vac6spur},
we show the corresponding vacuum energy density. Because the Hamiltonian 
in this case becomes non-hermitian, we call these solutions
spurious. The spurious solutions appear only after the inclusion of
$\phi^6$ interaction. Thus, they are consequences of the higher nonlinearity
in the OR equations. Since they don't seem to bear any interesting duality 
property that we discuss in this work, we don't present their results in 
the main text but just summarize those in the Appendix C with some 
discussion.

The most interesting feature of the $\phi^6$ theory is the appearence of
the two apparently different quasiparticle perturbation theories 
(see the upper branches of Figs. \ref{phasesnew1} and \ref{phasesnew2})
which represent the initial
nonperturbative theory. The first one corresponds to the symmetric
phase with the dimensionless quasiparticle couplings given by 
Eq.(\ref{pertunbr}) and the second one corresponds to the broken-symmetry 
phase with the dimensionless couplings given by Eq.(\ref{pertbr}).
The subscripts ``s'' and ``bs'' of the coupling $\chi$ and the mass $M$
represent the correspondence to the symmetric and broken-symmetry phases,
respectively. The two quasiparticle theories ({\it i.e.} $\chi_s$'s and 
$\chi_{bs}$'s) differ from each other significantly in the sence that
 one of them ($\chi_s$'s) preserves the 
$\phi\to -\phi$ symmetry of the initial Hamiltonian (\ref{clasphi6})
upon changing $\Phi\to -\Phi$, while 
the other ($\chi_{bs}$'s) does not respect the $\Phi\to -\Phi$ symmetry.
Physically, this means that in the theory of symmetric phase with 
the $\chi_s$ couplings the processes
involving odd number of particles are not allowed while they are 
allowed in the broken-symmetry theory with $\chi_{bs}$ couplings. This is 
because in the theory of symmetric phase,
the Hamiltonian depends on the even powers of $\Phi$
(to provide $\Phi \to -\Phi$ symmetry) and thus the S-matrix contains only 
the even number of creation and annihilation operators.

However, one may wonder if these two apparently different quasiparticle
theories are in fact equivalent to each other describing the same physics
with just different degrees of freedom. In this respect,
we notice that the two different quasiparticle theories give
the same effective potential in the limit of very large $G$ and 
$H$. In Figs.\ref{effpot1},\ref{effpot2} and \ref{effpot3}
we show the classical potentials which
correspond to the two quasiparticle theories at different values of 
the coupling $\chi_s^{(6)}$.
As it can be seen, when the quasiparticle coupling $\chi_s^{(6)}$ 
becomes small ({\it i.e.} $H$ becomes very large), the relative difference
between the two quasiparticle 
potentials diminishes substantially. We verified that the two classical
potentials are indeed identical to each other in the $\chi_s^{(6)}\to 0$
limit.

In the GEP approach \cite{gep6}, it was also recognized that there are three
solutions of symmetric phase (one of them is the trivial solution and two
others are the nontrivial solutions corresponding to the ones that we 
showed in Fig.\ref{phasesnew1}). As we consider only one
of the two nontrivial solutions (the upper branch in Fig.\ref{phasesnew1})
as the physically relevant (duality-related) solution, the GEP approach
also dealt with only one (physically relevant)
solution among the two nontrivial solutions.  
In Ref.\cite{gep6} it was shown that the form of $\bar \Omega$ equation
of the GEP approach is invariant under the exchange of $m$ and $M_s$
with an appropriate redefinition of effective couplings
$G'=(G-\frac{15H}{\pi}t_0)e^{-t_0}$ and $H'=He^{-t_0}$, as well as
$t'=t-t_0$, where $t_0=\ln\frac{M_s^2}{m^2}$ satisfies the 
constraint equation given by Eq.(\ref{equ2}): $1-e^{-t_0}=
\frac{3Gt_0}{4\pi}-\frac{45Ht_0^2}{8\pi^2}$ (Note $B=0$ for the symmetric
phase). Indeed, we can show the form
invariance of both OR Eqs. (\ref{equ1}) and (\ref{equ2}) under
the same transformations.

Observing such form invariance, it was argued in the GEP approach \cite{gep6}
that the nontrivial symmetric phase solution is just a duplication of the 
trivial solution and thus should be avoided. However, as we discussed above,
the duality between the trivial and nontrivial symmetric phase 
solutions provides the physical meaning to the form invariance of 
the OR equations. Because of the duality, we see the utility of 
both trivial and nontrivial solutions in the symmetric phase.
While the trivial 
solution can be utilized for the perturbation theory at small $G$ and $H$,
the nontrivial symmetric phase solution can be used for the construction
of quasiparticle perturbation theory in the regime of very large $G$ and $H$.

Furthemore, since there exists also a duality between the trivial 
solution and the nontrivial broken-symmetry solution we can rather easily 
understand the equivalence of the classical effective potentials
between the symmetric phase and the
broken-symmetry phase for the very large couplings of the original theory
(see Fig.\ref{effpot3}).
In the limit of $\chi_s^{(6)}\to 0$ (which means also $\chi_{bs}$'s$\to 0$,
as shown in Fig.\ref{couplings}),
the higher order corrections get negligible and thus the equivalence of 
the classical potentials between the symmetric and broken-symmetry phases 
is revealed manifestly as shown in Fig.\ref{effpot3}.

\section{Summary and Conclusions}
The formation of quasiparticle in RQFT may be a good example of 
self-organizing nature in the nonperturbative strong interactions. This 
nature is realized
by altering the vacuum structure and lowering the vacuum energy density
as the strength of the interactions gets more enhanced. Our main goal in this
paper was to provide the explicit examples that reveal such nature of
self-organizing RQFT and show the utility of duality generated by the
canonical transformation in solving the nonperturbative strong interaction
problem. For the explicit but simple examples, we analized
of 1+1 dimensional $\phi^4$ and $\phi^6$ theories. Our purpose was 
fulfilled by applying the OR method which is a generalization of the 
canonical quantization scheme.

The OR method uses canonical commutation relations for
the free field operators and employs
the canonical transformation of quantum fields represented by the change
of the field mass and the condensation of the field.
The idea to demand the right form of the Hamiltonian,
{\it i.e.} excluding the terms linear in $\Phi$ and
taking the coefficient of $\Phi^2$ as $M^2/2$ 
guarantees that the classical potential has the minimum at the
quantization point $\phi^{class}_M=b$ and that the double derivative of the
classical potential at this point is given by $M^2$. This assures that the OR 
method gives a self-consistent procedure of quantization and a 
valid description of the field excitations around the minimum point in 
terms of quasiparticles.

In contrast to the GEP and HA methods, the OR method is not based
on the variational procedure so that it does not have a bottleneck
of renormalization problem in higher dimensions as discussed in Section II.
However, the OR method has not been utilized as extensively as the GEP and HA 
methods. Even in the 1+1 dimensional scalar field theory, not much of the 
OR results were contrasted from the GEP and HA results although the OR 
results for the $(\phi^4)_{1+1}$ theory were presented in 
Ref.\cite{efbook}. We attempted to fill such gap in 
this work.

For the fixed classical
Hamiltonian and renormalization procedure (fixed by the counter-terms),
it is natural to expect that the OR method yields several different 
solutions for the quasiparticle mass and condensation because the OR 
equations are in general nonlinear. However, the unitary inequivalence
arising from these different OR vacuum solutions prohibits the 
transitions among the quasiparticles with different mass and/or condensation.
Although some of the solutions correspond
to the heavier quasiparticles, they do not decay into the light ones
due to the unitary inequivalence;
{\it i.e.} each sort of the quasiparticles is stable within the OR.
The unitary inequivalence between the OR solutions is shown in Appendix B.
A recent application of the unitary inequivalence to the flavor
mixing phenomena
can be found in Ref.\cite{ji-mish}. In this work, we selected only
the duality-related OR solutions because they can provide the immediate
physical interpretation. The reason why the physical interpretation
is immediate for the duality-related solutions is because the strong
coupling particle theory becomes solvable 
once it is transformed into an exactly equivalent quasiparticle theory 
with weak couplings.

\begin{figure}[htb]
\centering
\begin{minipage}[c]{0.45\hsize}
\epsfig{figure=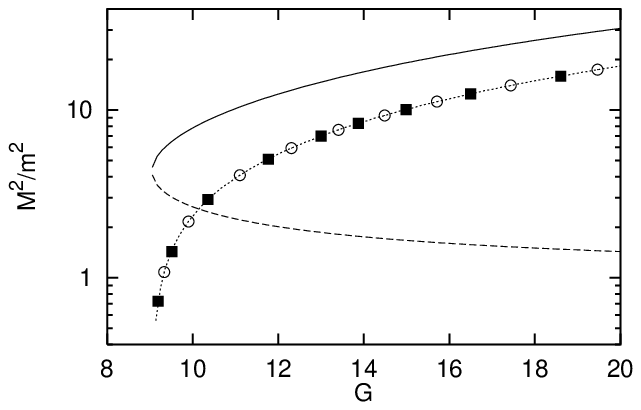, width=\hsize}
\caption{The nontrivial mass solutions for OR (solid and dashed
lines), GEP (dotted line) and Hartree methods (squares correspond to the
4$\pi c^2<3$ solution, circles to the 4$\pi c^2>3$ one). Here, $G=g/m^2$ and
for $G < 9.04$ there are no nontrivial solutions.
\label{geff}}
\end{minipage}
\hspace*{0.5cm}
\begin{minipage}[c]{0.45\hsize}
\vspace*{5mm}
\epsfig{figure=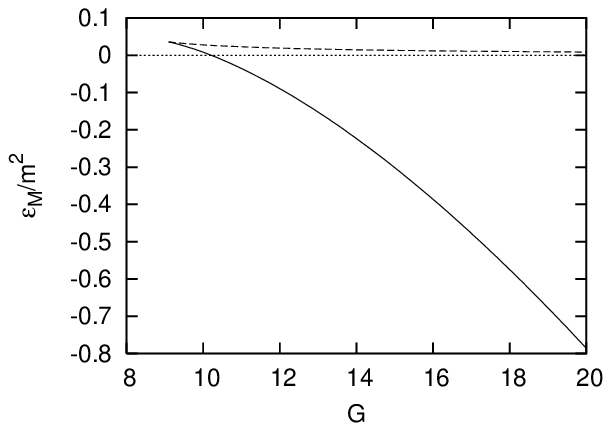, width=\hsize}
\caption{Vacuum energy densities of the nontrivial OR solutions
(solid and dashed
lines), GEP (dotted line) and Hartree methods (squares correspond to the
4$\pi c^2<3$ solution, circles to the 4$\pi c^2>3$ one). Here, $G=g/m^2$ and
for $G < 9.04$ there are no nontrivial solutions.
\label{energh0}}
\end{minipage}
\end{figure}

\begin{figure}[htb]
\centering
\epsfig{figure=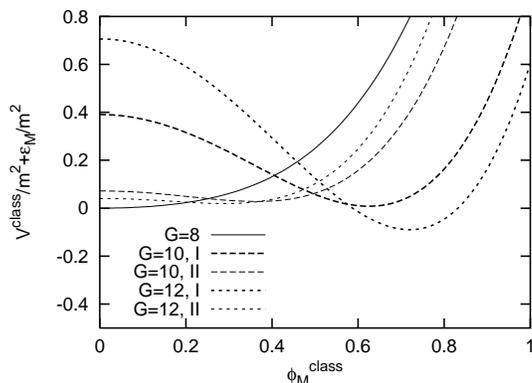,width=0.45\hsize}
\caption{The effective classical potential corresponding to 
Fig.~\protect\ref{geff} with $G =$ 8, 10 and 12. Although $G = 8$
has only a trivial effective potential denoted by a solid line,
both $G =$ 10 and 12 have two nontrivial effective potentials
denoted by I and II.\label{g4new}}
\end{figure}

Using the $(\phi^4)_{1+1}$ theory, we compared the OR results
with the results obtained by GEP and HA. In Ref.\cite{efbook}, it was 
discussed that the value of the critical coupling $G\approx 10.21$ 
obtained by the OR method
coincides with the one obtained by the GEP method.
We also note that this value of the critical coupling agrees 
remarkably well with the lattice result \cite{lattice}.
Due to the equivalence 
between the HA and GEP methods for the ground state energy \cite{CJT,ft7}
the agreement of the critical couplings between the OR and HA methods follows
as well. Although the critical values of coupling constant 
$G$ turned out to be same for all three methods (OR, GEP, HA),
the OR method gives
the quantitatively different dependence of the nontrivial field mass on the 
effective coupling constant, compare to the GEP and HA results
(see Fig.\ref{geff} and the discussions in Section II).

\begin{figure}[h]
\centering
\begin{minipage}[c]{0.45\hsize}
\epsfig{file=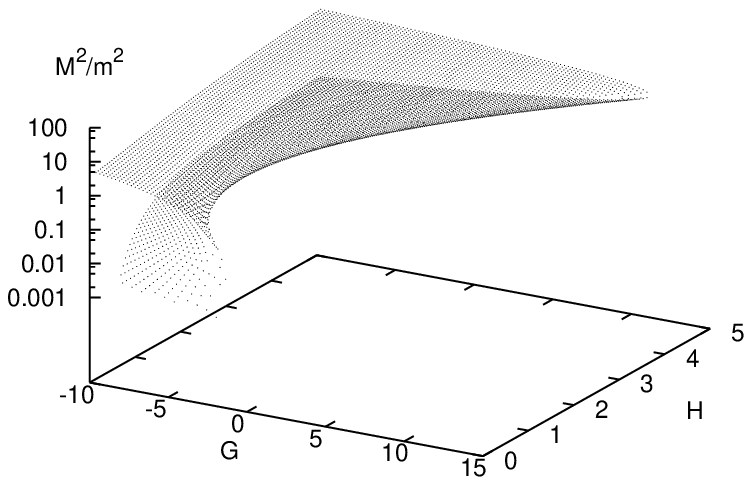,width=\hsize}
\caption{Nontrivial solutions of the OR equations for $\phi^4$ and
$\phi^6$ theory, symmetric phase.} 
\label{phasesnew1}
\end{minipage}
\hspace*{0.5cm}
\begin{minipage}[c]{0.45\hsize}
\epsfig{file=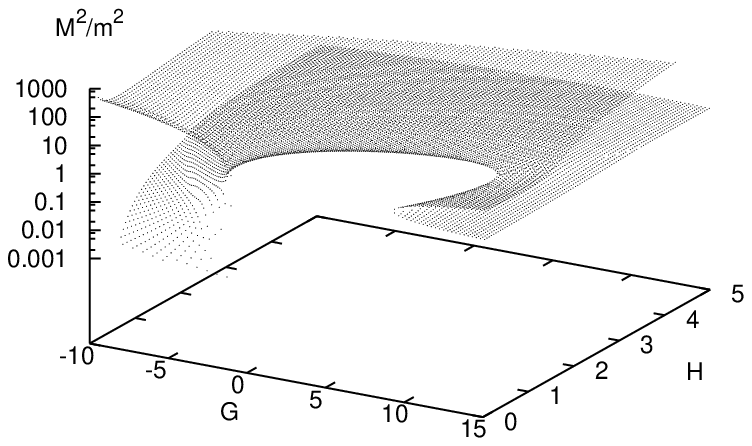,width=\hsize}
\caption{Nontrivial solutions of the OR equations for $\phi^4$ and
$\phi^6$ theory, broken-symmetry phase.} 
\label{phasesnew2} 
\end{minipage}
\end{figure}

We have extended the application to the 1+1 dimensional 
$\phi^6$ theory and
compared the results of OR method with what was already obtained 
by the GEP. We found that the different solutions for the mass of 
quasi-particles between the OR and the GEP are sustained.
In the OR analysis of $(\phi^6)_{1+1}$ theory we found two physically 
meaningful duality-related solutions (upper sheets of Figs.\ref{phasesnew1}
and \ref{phasesnew2}), one in the symmetric phase and the other in the 
broken-symmetry phase. Although these two solutions have apparently 
different symmetry property, we find that their effective potentials do 
agree in the limit of very large couplings $G$ and $H$ of the initial theory,
{\it i.e.} very small effective couplings $\chi_s$'s and $\chi_{bs}$'s of 
the quasiparticle theories. This indicates an existence of a unique 
effective 
potential to the quasiparticles regardless of their symmetry properties. 
The actual perturbation theory using the quasiparticles degrees of 
freedom with the systematic higher order corrections may deserve further 
investigation.

\begin{center}
{\large\bf Acknowledgments}
\end{center}
CRJ thanks Yuriy Mishchenko for many useful discussions.
This work is supported by the Brain Korea 21 project in year 2001, the Korea 
KRF 2001-015-DP0085, the Korea KOSEF 1999-2-111-005-5, the 
RFBR-01-02-16431, the RFBR-03-02-17291, the INTAS-2000-366 and the US DOE 
under grant No. DE-FG02-96ER40947. The North Carolina Supercomputing 
Center and the National Energy Research 
Scientific Computer Center are also acknowledged for the grant of Cray time.

\setcounter{equation}{0}
\renewcommand{\theequation}{\mbox{A\arabic{equation}}}
\begin{center}
{\bf APPENDIX A: Hartree Approximations in the $(\phi^4)_{1+1}$ Theory}
\end{center}
The essential idea of the Hartree approximation\cite{Chang} is to linearize
the non-linear field equations using the mean fields,
{\it e.g. $\phi^3 \rightarrow 3 \langle \phi^2 \rangle \phi -2
\langle \phi \rangle ^3$, etc.},
splitting the fields into classical mean fields
($\phi_c = \langle \phi \rangle$)
plus the quantum fields($\phi_q$), {\it i.e. $\phi = \phi_c + \phi_q$}.
In the classical limit, we know that the ground state is given by
$\phi^2=c^2$ if $c^2$ is defined by $c^2 = -\frac{m^2}{g}$ and positive
({\it i.e. $c^2>0$}).
The quantum fluctuations $\langle \phi_q^2(x) \rangle$ are found to be 
infinite.
However, one can renormalize
the theory using a mass counter term $\frac{1}{2} B \phi^2$, where
the constant $B$ is determined by requiring that $\phi_c = c$ is a static
solution (corresponding to the so-called ``abnormal vacuum state" or a
sort of vacuum condensate)
of a Hartree equation and this leads to
$B= -3 g \langle \phi_q^2 \rangle_{\phi_c =c}$.
After renormalization, the effective potential $V(\phi_c)$ is obtained in
terms of the subtracted quantum fluctuation, {\it i.e.
$\Delta \langle \phi_q^2 \rangle = \langle \phi_q^2 \rangle_{\phi_c}
- \langle \phi_q^2 \rangle_{\phi_c = c}$},
which satisfies the iterative (mass-gap type) equation\cite{Chang};
\begin{equation}
\Delta\langle\phi_q^2\rangle = \frac{1}{4\pi} \ln\frac{2c^2}
{3\phi_c^2 -c^2 + 3\Delta\langle\phi_q^2\rangle}.
\label{gapeq}
\end{equation}
Using the solutions of Eq.(\ref{gapeq}), one can find that $\phi_c = c$
is a local minimum (maximum) of $V(\phi_c)$ for $4\pi c^2 >3 (0<4\pi c^2<3)$
while $\phi_c=0$ is always a local minimum. For $0<4\pi c^2<3$, a nontrivial
local minimum exists at $\phi_c = \sqrt{c^2 -
3\Delta\langle\phi_q^2\rangle} = c^\prime$.
Expanding $V(\phi_c)$ in each of the local minima, one can introduce the
following mass parameters;
\begin{eqnarray}
m_0^2 = \frac{d^2 V(\phi_c)}{d\phi_c^2}\bigg|_{\phi_c=0}
= g(3\Delta\langle\phi_q^2\rangle|_{\phi_c=0} - c^2), \nonumber \\
m_c^2 = \frac{d^2 V(\phi_c)}{d\phi_c^2}\bigg|_{\phi_c=c}
= 2gc^2 \frac{8\pi c^2 -6}{8\pi c^2 +3}, \nonumber \\
m_{c^\prime}^2 = \frac{d^2 V(\phi_c)}{d\phi_c^2}\bigg|_{\phi_c=c^\prime}
= 4g(c^2-3\Delta\langle\phi_q^2\rangle)
\frac{4\pi(c^2-3\Delta\langle\phi_q^2\rangle)-3}
{8\pi(c^2-3\Delta\langle\phi_q^2\rangle)+3}.
\label{masses}
\end{eqnarray}
Then, the intrinsic strength measured in terms of these mass parameters
are given by the dimensionless couplings; {\it i.e.}
\begin{equation}
g_0 = \frac{g}{m_0^2}, \mbox{\hskip 5mm} g_c = \frac{g}{m_c^2},
\mbox{\hskip 5mm} g_{c^\prime} = \frac{g}{m_{c^\prime}^2}.
\label{dimlesscoupling}
\end{equation}
For the entire $c^2>0$ region, it turns out that the nontrivial vacuum
solutions exist only for $g_0 \geq g_{crit} \approx 9.045$ (note that it
is not the same as the critical coupling of the phase transition $G=10.21$
as discussed in Section II),
where the critical coupling $g_0 = g_{crit}$ occurs at $4\pi c^2 =3$.
Thus, it appears that one can find the two nontrivial vacuum solutions;
one for $4 \pi c^2 > 3$ and the other for $0< 4 \pi c^2 < 3$. However,
we find that the two solutions are indeed identical in the sense that
$g_c$ and $g_{c^\prime}$ coincide when we plot them (see 
Fig.~\protect\ref{app}) as a function of $g_0$,
{\it i.e.} $g_c(g_0) = g_{c^\prime}(g_0)$ for all $g_0 \geq g_{crit}$.

We also notice a duality between the theory with $g_0$ based on the
trivial vacuum $\phi_c =0$ and the theory with $g_c$ (or $g_{c^\prime}$)
based on the nontrivial vacuum $\phi_c = c$ for $4 \pi c^2 > 3$
(or $\phi_c = c^\prime$ for $0 < 4 \pi c^2 < 3$).
Namely, a strong coupling $g_0$ theory is identical to a weak coupling
$g_c$ (or $g_{c^\prime}$) theory.

\setcounter{equation}{0}
\renewcommand{\theequation}{\mbox{B\arabic{equation}}}
\begin{center}
{\bf APPENDIX B: Unitary Inequivalence between the Nontrivial OR
Solutions}
\end{center}

In this Appendix, we show the unitary inequivalence between a vacuum
with $b_1$ and $t_1$ and another vacuum $b_2$ and $t_2$, where $t_1\neq
t_2$ or $b_1 \neq b_2$.

The canonical transformation taking a trivial vacuum ({\it i.e.}
$b=0$ and $t=1$) into a nontrivial one ({\it i.e.} $b \neq 0$ and 
$t \neq 1$) has the following form ({\it e.g.} for the bosons)
\begin{equation}
\label{eq01z}
\begin{array}{cc}
1. & a_k \rightarrow a_k-2\pi mb\delta (k); \\
2. &
\begin{array}{c}
a_k \rightarrow a_k\cosh (\lambda )-a_{-k}^{\dagger}\sinh (\lambda ), \\
a_k^{\dagger} \rightarrow a_k^{\dagger}\cosh (\lambda )-a_{-k}\sinh
(\lambda ).
\end{array}
\end{array}
\end{equation}
Their corresponding operators have the form
\begin{equation}
\label{eq02z}
\begin{array}{c}
U_1=\exp \left\{ -2\pi mb(a_0-a_0^{\dagger })\right\};  \\
U_2=\exp \left\{ \frac 12\int d\vec k\lambda (k)(a_{-k}a_k-a_k^{\dagger
}a_{-k}^{\dagger })\right\}
\end{array}
\end{equation}
with $\lambda (k)=\frac 12\ln \frac{\omega (k)}{\omega (k,t)}$ and $\omega
(k)=\sqrt{k^2+t^2m^2}$.

To connect the two vacua, one with $b_1$ and $t_1$ and another with
$b_2$ and $t_2$, we can use the transformation given by Eq.(\ref{eq02z}) 
with $b=b_2-b_1$ and $\lambda (k)=\frac 12\ln 
\frac{\omega _1(k)}{\omega _2(k)}$ where $\omega _i(k)=\sqrt{k^2+t_i^2m^2}$. 
To show that, let us define $V_i:\left\langle 0,1\right\rangle
\rightarrow
\left\langle b_i,t_i\right\rangle $ by the transformation given by
Eq.(\ref{eq02z}). Then
$$
\left\langle b_1,t_1\right\rangle \rightarrow \left\langle
b_2,t_2\right\rangle =\left\langle b_1,t_1\right\rangle \rightarrow
\left\langle 0,1\right\rangle \rightarrow \left\langle b_2,t_2\right\rangle
$$
and therefore $V=V_2V_1^{-1}:\left\langle b_1,t_1\right\rangle \rightarrow
\left\langle b_2,t_2\right\rangle $. Straightforward algebraic
manipulation with Eq.($\ref{eq02z}$) then results in the unitary
inequivalence between
the vacuum with $\left\langle b_1,t_1\right\rangle$ and the vacuum
with $\left\langle b_2,t_2\right\rangle$.

In Ref.\cite{efbook}, it was mentioned that%
$$
\begin{array}{c}
U_1=e^{-\frac 12(2\pi mb)^2[a_0,a_0^{\dagger }]}:U_1:, \\
U_2=e^{-\frac V{(2\pi )^3}\int d\vec k\ln (\cosh [\lambda (k)])}:U_2:.
\end{array}
$$
where $[a_0,a_0^{\dagger }]=\delta ^{(3)}(0)$. Here, we can equate 
$\delta^{(3)}(0)$ to $\frac V{(2\pi )^3}$. Note also
that transformation $U_1$ acts only on the $k=0$ field operators, therefore%
$$
\begin{array}{c}
U_2\circ U_1=\left( \prod\limits_{k\neq 0}U_2(k)\right) \left( U_2(0)\circ
U_1(0)\right) , \\
:U_2\circ U_1:=\left( \prod\limits_{k\neq 0}:U_2(k):\right) :U_2(0)\circ
U_1(0):.
\end{array}
$$
Hence%
$$
U_2\circ U_1=e^{-\frac V{(2\pi )^3}\int\limits_{k\neq 0}d\vec k\ln (\cosh
[\lambda (k)])}\prod\limits_{k\neq 0}:U_2(k):\left( U_2(0)\circ
U_1(0)\right) .
$$
Then for the inner product of the vacuum $\left| 0\right\rangle _1$
associated with $\left\langle b_1,t_1\right\rangle $ and $\left|
0\right\rangle _2$ associated with $\left\langle b_2,t_2\right\rangle $ we
have%
$$
_1\left\langle 0|0\right\rangle _2=\prod {}_{1,k}\left\langle
0|0\right\rangle _{2,k}\leq \prod_{k\neq 0}{}_{1,k}\left\langle
0|0\right\rangle _{2,k}=e^{-\frac V{(2\pi )^3}\int\limits_{k\neq 0}d\vec
k\ln (\cosh [\lambda (k)])}\rightarrow 0,
$$
as $$\text{ }V\rightarrow \infty \text{.}$$
The two vacua therefore are unitary inequivalent if their $t$ parameters
are
different. If $t_1=t_2$ but $b_1\neq b_2$ then $\lambda (k)\equiv 0$ and $%
V=U_1(b_2-b_1).$ Then%
$$
_1\left\langle 0|0\right\rangle _2=\prod {}_{1,k}\left\langle
0|0\right\rangle _{2,k}={}_{1,k=0}\left\langle 0|0\right\rangle
_{2,k=0}=e^{-\frac 12(2\pi mb)^2\delta (0)}=0\text{,}
$$
in the infinite volume limit.

Therefore $_1\left\langle 0|0\right\rangle _2=0$ in the infinite volume
limit whenever $t_1\neq t_2$ or $b_1\neq b_2$.

\begin{figure}[htb]
\centering
\epsfig{file=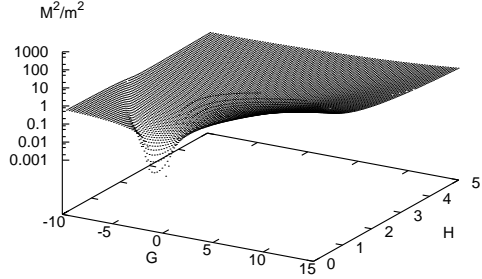,width=0.45\hsize}
\caption{Nontrivial solutions of the OR equations for $\phi^4$ and
$\phi^6$ theory, imaginary $b$ phase.} 
\label{phasesnew3} 
\end{figure}

\begin{figure}[htb]
\centering
\begin{minipage}[c]{0.45\hsize}
\epsfig{file=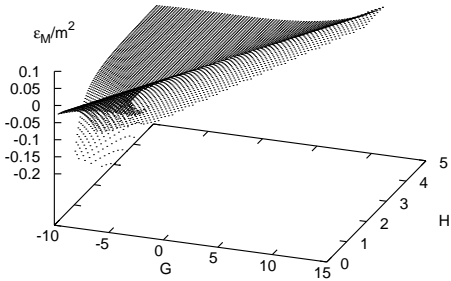,width=\hsize}
\end{minipage}
\hspace*{0.5cm}
\begin{minipage}[c]{0.45\hsize}
\epsfig{file=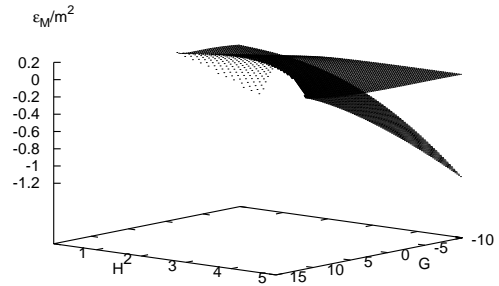,width=\hsize}
\end{minipage}
\caption{Vacuum energy density of nontrivial solutions
of the OR equations for $\phi^4$ and $\phi^6$ theory, symmetric phase.
The two plots (right and left) are identical but presented in two different
angles of view. The $90^{\circ}$ clockwise rotation of the left plot
around the axis of $\varepsilon_M/m^2$ is identical to the right plot.} 
\label{vac6sym}
\end{figure}

\begin{figure}[h]
\centering
\begin{minipage}[c]{0.45\hsize}
\epsfig{file=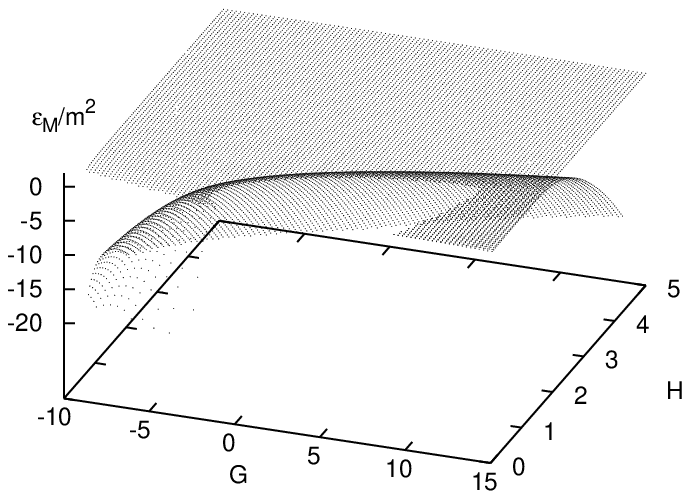,width=\hsize}
\caption{Vacuum energy density of nontrivial solutions
of the OR equations for $\phi^4$ and
$\phi^6$ theory, broken-symmetry phase.} 
\label{vac6brok} 
\end{minipage}
\hspace*{0.5cm}
\begin{minipage}[c]{0.45\hsize}
\epsfig{file=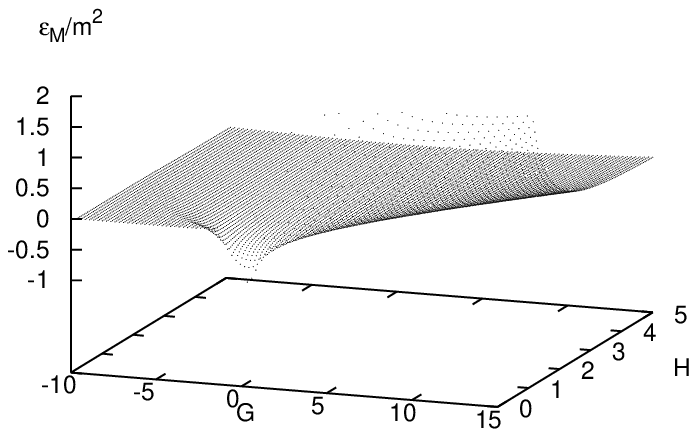,width=\hsize}
\caption{Vacuum energy density of nontrivial solutions
of the OR equations for $\phi^4$ and
$\phi^6$ theory, imaginary $b$ phase.} 
\label{vac6spur} 
\end{minipage}
\end{figure}

\begin{figure}[h]
\centering
\begin{minipage}[c]{0.45\hsize}
\epsfig{file=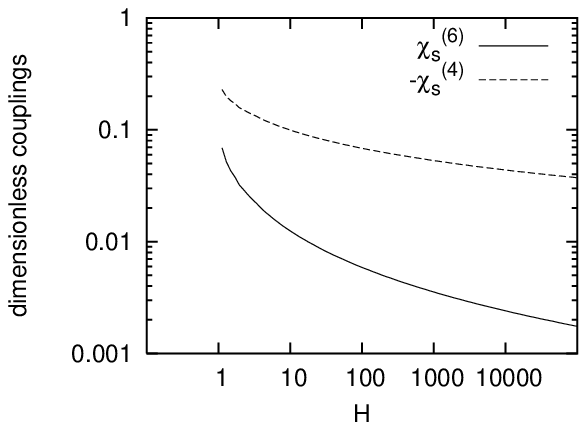,width=\hsize}
\caption{Dimensionless couplings of the symmetric perturbative
solution for $G=0$.} 
\label{coup1}
\end{minipage}
\hspace*{0.5cm}
\begin{minipage}[c]{0.45\hsize}
\epsfig{file=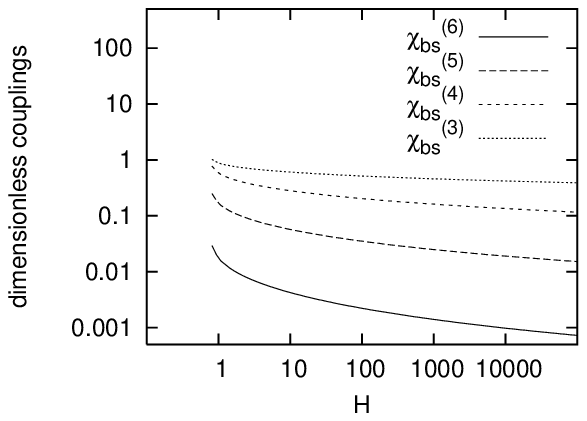,width=\hsize}
\caption{Dimensionless couplings of the broken-symmetry perturbative
solution for $G=0$.} 
\label{coup2} 
\end{minipage}
\end{figure}

\begin{figure}[h]
\centering
\begin{minipage}[c]{0.45\hsize}
\epsfig{file=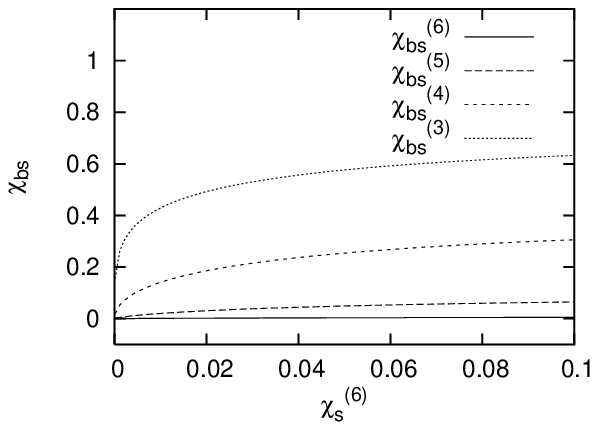,width=\hsize}
\caption{Dimensionless couplings of the broken-symmetry perturbative
solution versus the dimensionless coupling of the symmetric
solution ($G=0$).} 
\label{couplings}
\end{minipage}
\hspace*{0.5cm}
\begin{minipage}[c]{0.45\hsize}
\epsfig{file=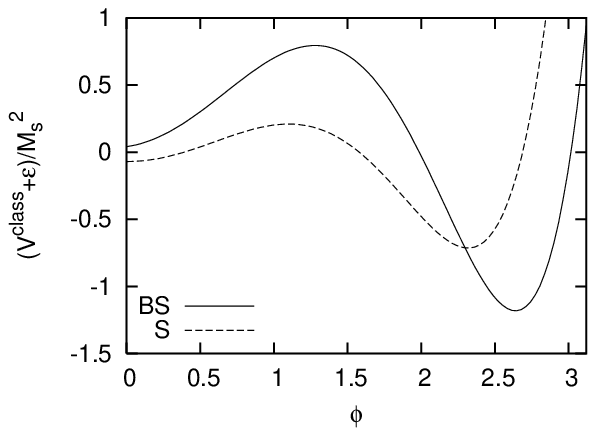,width=\hsize}
\caption{Classical potential of broken-symmetry (BS) and symmetric (S) 
theories at $\chi^{(6)}_s=5.85\cdot 10^{-3}$ ($H=100, G=0$).}
\label{effpot1} 
\end{minipage}
\end{figure}

\begin{figure}[h]
\centering
\begin{minipage}[c]{0.45\hsize}
\epsfig{file=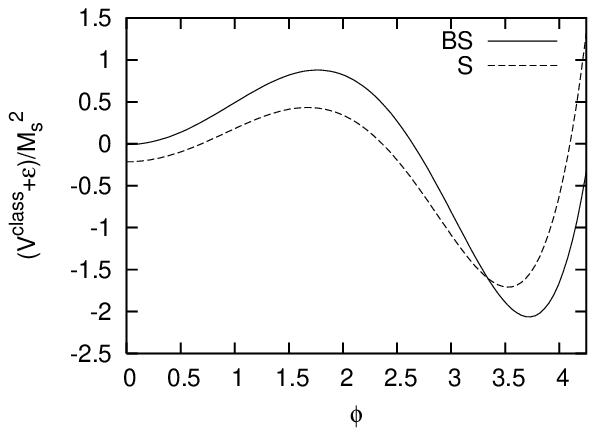,width=\hsize}
\caption{Classical potential of broken-symmetry (BS) and symmetric (S) 
theories at $\chi^{(6)}_s=1.33\cdot 10^{-3}$ ($H=10^6, G=0$).}
\label{effpot2}
\end{minipage}
\hspace*{0.5cm}
\begin{minipage}[c]{0.45\hsize}
\epsfig{file=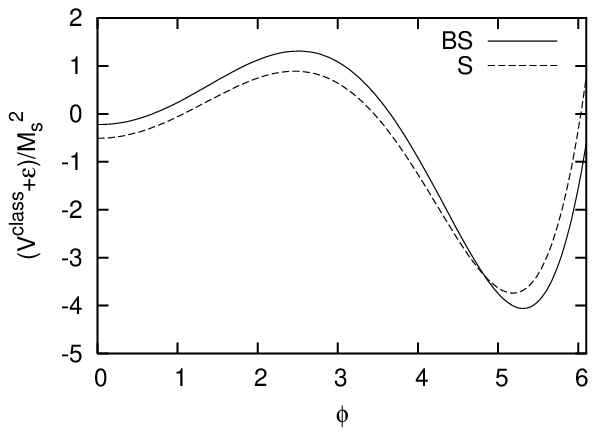,width=\hsize}
\caption{Classical potential of broken-symmetry (BS) and symmetric (S) 
theories at $\chi^{(6)}_s=1.02\cdot 10^{-3}$ ($H=10^{15}, G=0$).}
\label{effpot3} 
\end{minipage}
\end{figure}

\begin{figure}[htb]
\centering
\epsfig{file=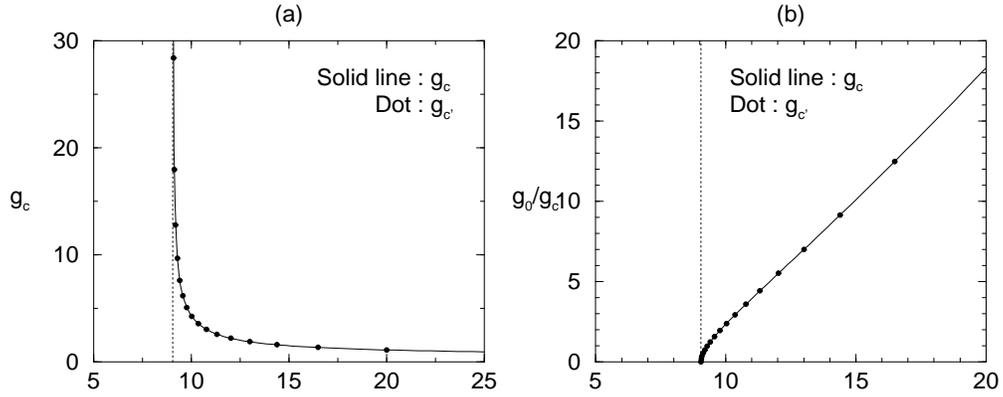, width=0.8\hsize, angle=270}
\caption{(a) $x$ axis is $g_0$ and $y$ axis is $g_c$ (solid line)
and $g_{c'}$ (dot). (b) $x$ axis is $g_0$ and $y$ axis is
$\frac{g_0}{g_c}$ (solid line) and $\frac{g_0}{g_{c'}}$ (dot).}
\label{app}
\end{figure}

\setcounter{equation}{0}
\renewcommand{\theequation}{\mbox{B\arabic{equation}}}
\begin{center}
{\bf APPENDIX C: Spurious Solutions in the $(\phi^6)_{1+1}$ Theory}
\end{center}
Among the solutions of OR Eqs.(\ref{equ1}) and (\ref{equ2}) for 1+1
dimensional $\phi^6$ theory there is one with negative $B$ that gives
pure imaginary condensate $b=i\sqrt{|B|}$. This solution is shown
in Fig.\ref{phasesnew3} and the corresponding vacuum energy density is shown
in Fig.\ref{vac6spur}.
In this case the Hamiltonian becomes non-hermitian because
it contains anti-hermitian terms in odd powers of $\Phi_M$.
When one calculates average values
of such operators in any state with definite number
of particles, these terms drop out, because odd number of creation
and annihilation operators can not leave the number of particles 
unchanged. However, the physical states are complicate and 
contain indefinite number of particles
so that the antihermitian terms would contribute to the physical
observables and give them imaginary expectation values.
Moreover, our starting Hamiltonian density 
given by Eq.(\ref{clasphi6}) involves only the real (not complex as in 
the O(2) symmetric theory) scalar field $\phi_m$ and even after shifting 
the quantization point the field $\Phi_M$ is required to remain as real
not complex. For these reasons, we call the imaginary $b$ solution 
spurious. In principle, we can throw away the spurious solution and
do not consider the imaginary-$b$ phase in the simple one-component model.

Nevertheless, in the history of many-body analysis sometimes the occurance
of such spurious solution (forbidden in principle) led to interesting
physics. For example, the occurance of imaginary solution has been 
previously observed in the analysis of many-body system when the method 
gets more accurate but the number of degrees of freedom is not sufficient.
Such example may be illustrated in the random phase approximation (RPA) 
calculation of a fermion system\cite{Kadanoff} where the imaginary energy 
poles are found if the fermion interaction is attractive and weak below a 
certain critical temperature. These imaginary poles were 
interpreted\cite{Kadanoff} as the mathematical manifestation 
of the instability which leads to the superconducting state. Such 
imaginary poles were however not found in the less accurate method such as 
the Tamm-Dancoff approximation (TDA). 
Thus, if there is any possibility that the imaginary solution could 
actually mean something interesting in the many-body analysis,
we should not overlook such a possibility.

Due to such a tantalizing possibility that the imaginary solution could 
occur indicating an existence of a new phase, we note a few more 
particular features of the spurious solutions 
instead of simply throwing those solutions away. 
In particular, it may be interesting to note that the corresponding 
quasi-particle mass values become quite small ($M<<m$) 
when the imaginary $b$ solutions have the least vacuum energy density
as shown in Figs.\ref{phasesnew3} and \ref{vac6spur}.
We do not know yet, however, whether these spurious solutions are anything
to do with the condensation of (zero-mass) Goldstone-bosons if the degrees 
of freedom are added in such a way that the fields become complex and a 
continuous symmetry (see however the discussion for O(2) symmetry near the 
end of this Appendix) is restored.  
With the limited degrees of freedom at hands, the theory we 
consider in the present work does not lead to an appearance
of some effective charge in the theory. For the spurious solutions, 
the point of quantization is fixed as pure imaginary and can not be 
rotated by any arbitrary phase.
Also, in the effective Hamiltonian 
given by Eq.(\ref{rightH}), $b$ can be thought as a part of effective
coupling. For example, consider the term $g b \Phi^3_M$ in 
Eq.(\ref{rightH}) and build up an effective potential between two $\Phi_M$ 
particles intermediating another $\Phi_M$ as an exchange-boson. Then, the 
coefficient $g b$ of $\Phi^3_M$ is an effective coupling of the triple 
vertices and the square of the coefficient would determine an overall sign 
of effective potential. The detailed form of the potential is of course 
determined only after taking into account the propagator of 
exchange-boson. However, the overall sign of effective potential 
determines whether the force between the two $\Phi_M$ particles is 
attractive or repulsive. Due to the sign change of the potential, if some 
bound states may not be formed (although they could be formed for the 
real $b$),
then the corresponding energy eigenvalues would appear as complex values 
indicating the instability of those states. In some sense, the imaginary
$b$ resembles a sort of optical potential which has been used rather 
frequently in the phenomenology of many-body nuclear physics.
A recent work showing an example of how the imaginary coupling can be 
used in practice was presented in Ref.\cite{neuberg}.

\begin{figure}[hbt]
\centering
\epsfig{file=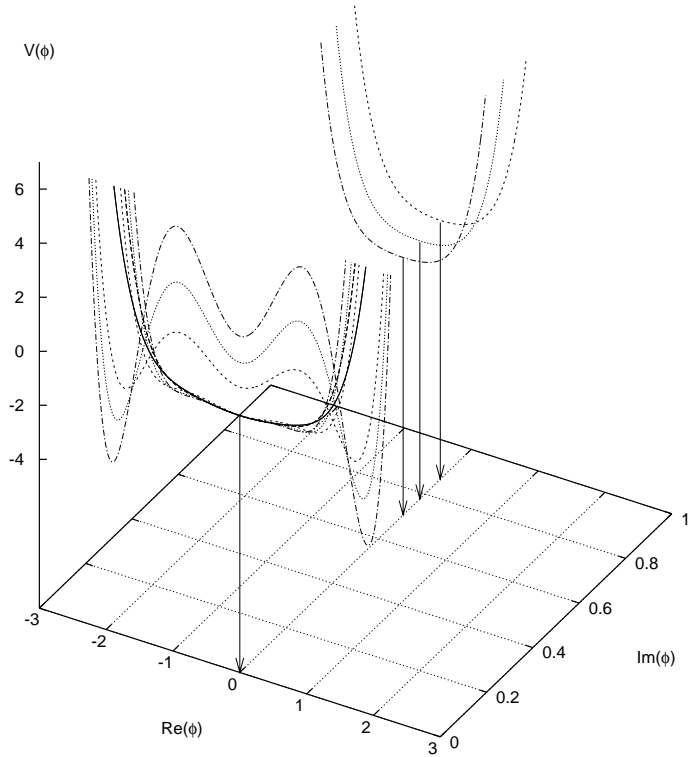,width=0.7\hsize}
\caption{ The dimensionless effective potential $V(\phi) = 
\Bigl[V^{class}+\varepsilon_M \Bigr]/m^2$ for $H/G = 
-0.209$($\alpha=-\beta$ in 
notations of Ref.\protect\cite{gep6}), where $\phi$ means 
$\phi^{class}_M$. $x$ axis is $Re(\phi)$,
$y$ axis is $Im(\phi)$ and $z$ axis is $V(\phi)$. In general, for
the complex $\phi$ the potential $V(\phi)$ is complex.
For $G=$-0.838(solid line) and -1.68(long-dashed line),
we show only the trivial solutions ($t=0$ and $b=0$) for the comparison with 
the 
nontrivial solutions for $G=$ -2.51(short-dashed line),-3.35(dotted line)
and -4.19(dot-dashed line). 
For each coupling constant, $G=$ -2.51(short-dashed line), -3.35(dotted 
line) and -4.19(dot-dashed line), there are two real $b$ 
broken-symmetry solutions
and one pure imaginary $b$ solution. The two real $b$ 
broken-symmetry solutions for each
coupling constant($G=$-2.51,-3.35 and -4.19) are shown on the $Im(\phi)=0$ 
plane. For the pure imaginary $b$ solutions,
the potential graphs are drawn
on the planes off from the $Im(\phi)=0$ plane with the different 
imaginary $b$ values depending on the $G$ values.} 
\label{overallV6}
\end{figure}

To show an overall picture of the effective potential in $(\phi^6)_{1+1}$ 
theory, we present the dimensionless effective potential $V(\phi) = \Bigl 
[V^{class}(\phi^{class}_M) + \varepsilon_M 
\Bigr ] /m^2$ for the broken-symmetry solution and the spurious solution
with $H/G = -0.209$($\alpha=-\beta$ in notations of 
Ref.\protect\cite{gep6}) in Fig.\ref{overallV6} for several 
different $G$ values. 
For simplicity, we didn't plot $V(\phi)$ for the symmetric phase $(b=0)$
solution. 
In Fig.\ref{overallV6}, $\phi$ denotes 
$\phi^{class}_M$ and the symmetry $V^{class}(\phi^{class}_M) = 
V^{class}(-\phi^{class}_M)$ is manifest for the OR solutions
with $B = b^2 > 0$, {\it i.e.} $Im(\phi) = 0$. 
Indeed, no matter what the sign of $B$ is, we can in general prove the 
symmetry $V(\phi) = V(-\phi)$ using the OR equations given by 
Eqs.(\ref{OR1}) and (\ref{OR2}). 
Thus, for the spurious solutions drawn off from $Im(\phi) = 0$ plane,
one can realize the symmetry $V(\phi) = V(-\phi)$ by considering
both planes of $Im(\phi) = \pm |b|$ although only the $Im(\phi) = +|b|$
solutions are shown in Fig.\ref{overallV6}.
The solid, long-dashed, short-dashed, dotted and
dot-dashed lines stand for $G$=-0.838,-1.68,-2.51,-3.35 and -4.19 
($\alpha$=-0.1,-0.2,-0.3,-0.4 and -0.5 in 
notations of Ref.\protect\cite{gep6}), respectively.
For the coupling constants $G = $ -2.51 (short-dashed), -3.35 (dotted)
and -4.19 (dot-dashed), the effective potentials corresponding to the
three nontrivial solutions including
a spurious solution with the imaginary $b$ 
are shown in Fig.\ref{overallV6}:
{\it e.g.} three dot-dashed lines (the spurious one is on the plane 
$Im(\phi) \neq 0$ and the other two real broken-symmetry solutions are on 
the plane 
$Im(\phi) = 0$) are shown for the single value of $G= -4.19$.
For the comparison with the trivial solutions, we added the solid line
for $G= -0.838$ and the long-dashed line for $G= -1.68$ which have the 
minimum only at $Re(\phi) = Im(\phi) =0$. 

\begin{figure}[hbt]
\centering
\begin{minipage}[c]{0.45\hsize}
\epsfig{file=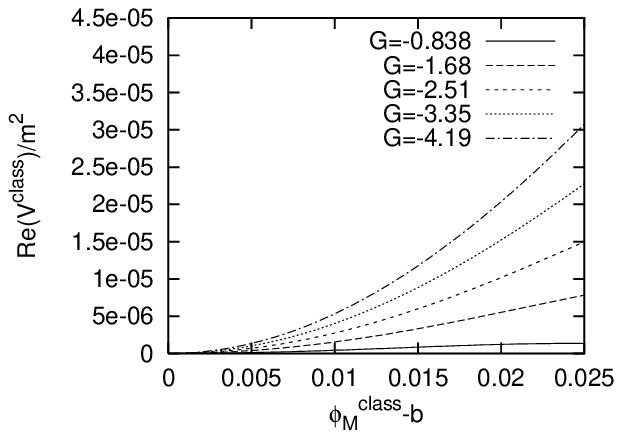,width=\hsize}
\caption{ Real part of effective potentials for the spurious solutions
of G=-0.838(solid line), -1.68(long-dashed line), -2.51(short-dashed 
line), -3.35(dotted line) and -4.19(dot-dashed line).}
\label{realV6III}
\end{minipage}
\hspace*{0.5cm}
\begin{minipage}[c]{0.45\hsize}
\epsfig{file=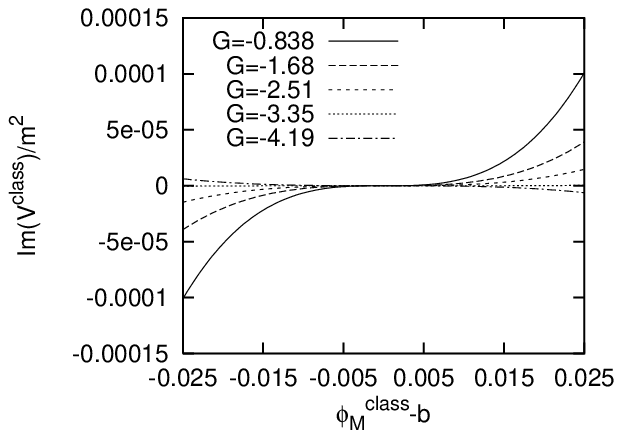,width=\hsize}
\caption{ Imaginary part of effective potentials for the spurious solutions
of G=-0.838(solid line), -1.68(long-dashed line), -2.51(short-dashed 
line), -3.35(dotted line) and -4.19(dot-dashed line).}
\label{imV6III}
\end{minipage}
\end{figure}

\begin{figure}[hbt]
\centering
\epsfig{file=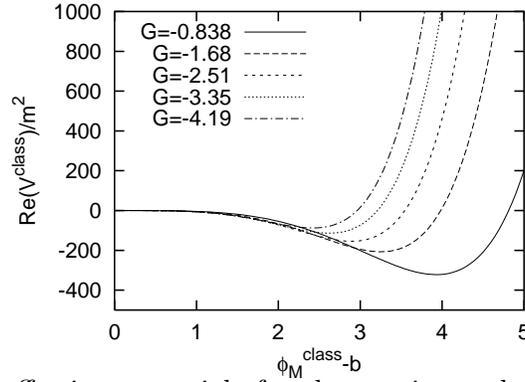,width=0.45\hsize}
\caption{ Real part of effective potentials for the spurious solutions
of G=-0.838(solid line), -1.68(long-dashed line), -2.51(short-dashed 
line), -3.35(dotted line) and -4.19(dot-dashed line) in the larger scale
of classical fields. The blow-up of the results in a tiny region of 
$\phi^{class}_M -b < 0.025 $ is shown in Fig.\protect\ref{realV6III}.}
\label{largeV6III}
\end{figure}

In Figs.\ref{realV6III}-\ref{largeV6III}, the effective potentials
for the spurious solutions are shown. Since the spurious solutions
have $Im(\phi) \neq 0$, the corresponding effective potentials $V(\phi)$ 
are complex. The real parts of effective potentials for the 
spurious solutions of $G=$-0.838(solid line), -1.68(long-dashed line), 
-2.51(short-dashed line), -3.35(dotted line) and -4.19(dot-dashed line)
are shown in Figs.\ref{realV6III} and \ref{largeV6III} while the imaginary
parts for the same coupling constants are shown in Fig.\ref{imV6III}.
Also, the horizontal axes of Figs.\ref{realV6III}-\ref{largeV6III} are 
given by the real field $\Phi^{class}_M = \phi^{class}_M -b$ rather than 
the complex field $\phi^{class}_M$ due to the imaginary $b$.
However, the effective potentials for $G=$ -2.51, -3.35 and -4.19 in 
Figs.\ref{realV6III}-\ref{largeV6III} correspond to the ones
drawn off from the plane $Im(\phi)=0$ in Fig.\ref{overallV6}.
As shown in Fig.\ref{vac6spur}, the energy density of the spurious 
solution gets the lowest among all the OR solutions
for the small coupling such as $G=$ -0.838 and -1.68. Especially, 
the quasi-particle mass gets close to zero as $G$ becomes close to zero.
This behavior may be noticed from Fig.\ref{realV6III} where the second 
derivative of $Re(V^{class})$ near $\phi^{class}_M = b$ is the smallest
for $G=$ -0.838. The corresponding behaviors of $Im(V^{class})$ are
shown in Fig.\ref{imV6III}. It is manifest from Fig.\ref{imV6III} that 
$Im(V^{class})$ contains only the odd powers of the real field 
$\Phi^{class}_M (= \phi^{class}_M -b)$, which can also be explicitly shown 
from our effective Hamiltonian given by Eq.(\ref{rightH}).
The large field behaviors of $Re(V^{class})$ are shown in 
Fig.\ref{largeV6III}. We can see that the minima of $Re(V^{class})$
appear in the large real field $\Phi^{class}_M$ for the spurious 
solutions, which one can also notice from the effective Hamiltonian 
(Eq.(\ref{rightH})) by keeping only the even powers of $\Phi_M$. 

\begin{figure}[htb]
\centering
\begin{minipage}[c]{0.45\hsize}
\epsfig{file=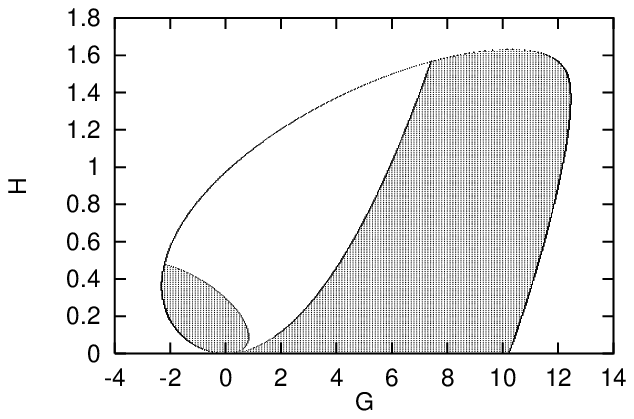,width=\hsize}
\caption{The domain of trivial solutions in $G$, $H$ parameter space
(shadowed area). If the spurious solution shown in 
Fig.\protect\ref{phasesnew3} is not taken into account, then the domain 
enlarges to the inside of the 
entire area bounded by the external solid line, which is equivalent to 
the GEP result shown in Fig.1 of Ref.\protect\cite{gep6}.} 
\label{rabbit} 
\end{minipage}
\hspace*{0.5cm}
\begin{minipage}[c]{0.45\hsize}
\epsfig{file=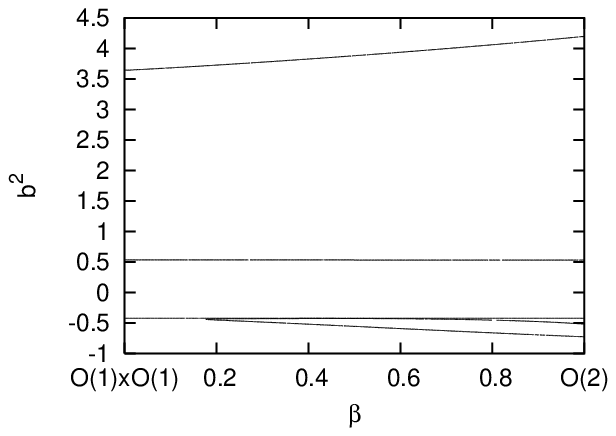,width=\hsize}
\caption{Correspondence between solutions of OR equations for O(1)$\times $O(1)
and O(2) $(\phi^6)_{1+1}$ theories at $G=1.0$, $H=3.0$.}
\label{bevol}
\end{minipage}
\end{figure}

In Fig.\ref{rabbit}, we show the domain of nontrivial solutions
that have the vacuum energy density less than zero
in the coupling space of $G$ and $H$. 
Without the spurious solutions, the domain of nontrivial OR solutions 
coincides with the GEP result (Fig.1 in Ref.\cite{gep6}) 
\footnote{In Ref.\cite{gep6}, Stevenson 
and Roditi used somewhat different
notations for couplings. Their couplings $\alpha$ and $\beta$ are related to 
our $G$ and $H$ as
$\alpha =\frac{3G}{8\pi}$ and $\beta=\frac{45H}{8\pi^2}$, respectively.
The boundary of allowed $\alpha$ and $\beta$ regions for the GEP 
were found as
a curve (see Fig.1 of Ref.\cite{gep6}) that can be drawn by the 
following 
formulas: $\alpha = (2z+3+(z-3)e^z)/z^2, \beta = 3(z+2+(z-2)e^z)/z^3$ with
$0<z<\infty$.} denoted by
the solid line of boundary. Taking into account the spurious solutions, 
however, the domain of trivial solutions reduces to the 
shaded region of the phase diagram shown in Fig.\ref{rabbit}.
The appearance of the spurious phase ($B = b^2 < 0$) in the domain where 
couplings are small may be interesting because it is quite opposite to the 
appearance of the normal nontrivial phase ($B > 0$), for which one would 
expect that the interaction term that leads to the 
formation of vacuum condensate is dominant at the larger coupling and 
therefore the nontrivial phases appear in the domain of large couplings. 
The presence of the nontrivial solution for any values of the couplings is
also interesting in connection with the
Haag's theorem \cite{haag}, according to which, if the vacuum 
state of a Lorentz invariant theory is unique, then all of its 
observables are identical with those of the free theory. 
Even very small admixture of $\phi^6$ interaction
to the $\phi^4$ theory allows to have a set of vacuum states for any
value of the $\phi^4$ coupling and thus to overcome the conditions of the 
theorem. 

In order to explore whether the spurious solution can be considered as
an evidence of some real solution by adding more degrees of freedom, 
we applied the OR method
to the two-component $(\phi^6)_{1+1}$ theory.
In this model, the imaginary condensation of one component could be 
represeneted by a real condensation of the other component.
However, we found that the imaginary-$b$ solution of the one-component
model corresponds to imaginary-$b$ solution of two-component O(2) model,
but not to the real solution.
The matching was done by solving an arbitrary two-component model
\begin{eqnarray}
\label{hamo2}
{\cal H}=N_m\Bigl ( \frac{1}{2}(\pi^2_1+\pi_2^2)+\frac{1}{2}((\nabla\phi_1)^2
+(\nabla\phi_2)^2)
+\frac{m^2}{2}(\phi_1^2+\phi_2^2)+ \\
\frac{g}{4}(\phi^4_1+\phi^4_2) + h(\phi^6_1+\phi^6_2) +
\beta(\frac{g}{2}\phi_1^2\phi_2^2 + 3h\phi_1^4\phi_2^2 + 3h\phi_1^2\phi_2^4)
\Bigr ). \nonumber
\end{eqnarray}
For $\beta=0$ and 1, Eq.(\ref{hamo2}) corresponds to O(1)$\times$O(1) and 
O(2) 
models, 
respectively. The evolution of $b^2$ for a ``typical''
value of $G$ and $H$ (we took $G=1, H=3$) is shown on Fig.\ref{bevol}.
As it can be seen, the negative $b^2$ solution in O(1)$\times$O(1) never
becomes positive as $\beta$ gets increased and even more spurious 
solutions appear in the limit of O(2). However, there is one-to-one 
correspondence between positive $b^2$ solutions in O(1)$\times$O(1) and 
O(2). We found that the same is true for any other $G$ and $H$.
Thus, our most immediate exploration didn't realize the anticipated
possibility that the negative $b^2$ solution in O(1)$\times$O(1) theory 
corresponds to a positive $b^2$ solution in O(2) theory. This may be a 
symptom of nonexistence of Goldstone 
mode in $1+1$ dimension known as the Coleman's theorem\cite{coleman1}.
Thus, further exploration including the higher dimension may be necessary
for the better understanding of the spurious solution.

\newpage

\begin{thebibliography}{99}
\bibitem{CJT}
J.M.Cornwall, R.Jackiw and E.Tomboulis, Phys. Rev. D{\bf 10}, 2428 (1974).

\bibitem{efimov}
G.~V.~Efimov,
Int.\ J.\ Mod.\ Phys.\ A {\bf 4}, 4977 (1989).

\bibitem{chang2}
S.~J.~Chang,
Phys.\ Rev.\ D {\bf 13}, 2778 (1976)
[Erratum-ibid.\ D {\bf 16}, 1979 (1976)].

\bibitem{magruder}
S.~F.~Magruder,
Phys.\ Rev.\ D {\bf 14}, 1602 (1976).

\bibitem{efbook}
M.~Dineykhan, G.~V.~Efimov, G.~Ganbold and S.~N.~Nedelko,
Lect.\ Notes Phys.\  {\bf M26}, 1 (1995).

\bibitem{ft7}
P.M. Stevenson, Phys. Rev. D {\bf 30}, 1712 (1984). 

\bibitem{gep}
P.M. Stevenson, Phys. Rev. D {\bf 35}, 2407 (1987).

\bibitem{gep4}
P.M. Stevenson, Phys.\ Rev.\ D {\bf 32}, 1389 (1985).

\bibitem{gep6}
P.~M.~Stevenson and I.~Roditi,
Phys.\ Rev.\ D {\bf 33}, 2305 (1986).

\bibitem{mp}
H. Mishra and A. R. Panda, J. Phys. G : Nucl. Part. Phys. {\bf 18}, 1301 (1992).

\bibitem{Chang}
S.~J.~Chang, Phys.\ Rev.\ D {\bf 12}, 1071 (1975).

\bibitem{hartree}
S.J.Chang and J.A.Wright, Phys. 
Rev. D {\bf 12}, 1595 (1975); S.J.Chang and T.-M.Yan, Phys. Rev. 
D {\bf 12}, 3225 (1975).  

\bibitem{coleman}
S.~R.~Coleman,
Phys.\ Rev.\ D {\bf 11}, 2088 (1975).

\bibitem{lattice}
W.~Loinaz and R.~S.~Willey,
Phys.\ Rev.\ D {\bf 58}, 076003 (1998)
[arXiv:hep-lat/9712008].

\bibitem{simon}
B.Simon, {\it The $P(\phi)_2$ Euclidean (Quantum) Field Theory},
Princeton University Press, New Jersey, 1974.

\bibitem{Munoz}
R.~Munoz-Tapia, J.~Taron and R.~Tarrach,
Int.\ J.\ Mod.\ Phys.\ A {\bf 3}, 2143 (1988).

\bibitem{wudka}
U.~Ritschel,
Z.\ Phys.\ C {\bf 54}, 297 (1992), \\
J.~Wudka,
Phys.\ Rev.\ D {\bf 37}, 1464 (1988).

\bibitem{ji-mish} C.-R.Ji and Y.Mishchenko, Phys. Rev. D {\bf 65}, 096015 
(2002); Phys. Rev. D {\bf 64}, 076004 (2001).

\bibitem{Kadanoff}
L.P.Kadanoff and P.C.Martin, Phys. Rev. {\bf 124}, 670 (1961);
D.J.Thouless, ``The Quantum Mechanics of Many-Body Systems", Second Edition,
Academic Press, New York and London, 1972.

\bibitem{neuberg}
N.E.Lighterink and B.L.G.Bakker, hep-ph/0010167.

\bibitem{haag}
R.~Haag, Dan.\ Mat.\ Fys.\ Medd. {\bf 29}, No.\ 12, 1 (1955), \\
A.~S.~Wightman and S.~S.~Schweber, Phys.\ Rev. {\bf 98}, 812 (1955).

\bibitem{coleman1}
S.~R.~Coleman,
Commun.\ Math.\ Phys.\  {\bf 31}, 259 (1973).
\end{thebibliography}
\end{document}